\documentclass[%
reprint,
amsmath,amssymb,
prb,
]{revtex4-2}
\usepackage[title]{appendix}
\usepackage{graphicx}
\usepackage{dcolumn}
\usepackage{bm}
\usepackage{gensymb}
\usepackage{float}
\usepackage{multirow}
\usepackage{tabularx}
\usepackage{amsmath}
\usepackage{color,soul}
\usepackage{makecell}
\usepackage{subcaption}
\usepackage{diagbox}
\usepackage{physics}
\usepackage[colorlinks,allcolors=black,pdftex]{hyperref} 
\usepackage[capitalise]{cleveref}
\usepackage[dvipsnames]{xcolor}

\renewcommand{\b}[1]{{\boldsymbol{#1}}}

\begin{document}

\preprint{APS/123-QED}


\title{Structured light and induced vorticity in superconductors II:  Quantum Print with Laguerre-Gaussian beam}

\author{Tien-Tien Yeh$^{1}$, Hennadii Yerzhakov$^{1}$, Logan Bishop-Van Horn$^{2}$, 
Srinivas Raghu$^{2}$,
Alexander Balatsky$^{1,3*}$}
\affiliation{$^{1}$Nordita, Stockholm University, and KTH Royal Institute of Technology, Hannes Alfvéns väg 12, SE-106 91 Stockholm, Sweden}
\affiliation{$^{2}$Department of Physics, Stanford University, Stanford, CA, USA}
\affiliation{$^{3}$Department of Physics, University of Connecticut, Storrs, Connecticut 06269, USA}

\begin{abstract}
Challenge to  control the quantum states of matter via light have been at the forefront of modern research on driven  quantum matter.
We explore the imprinting effects of structured light on superconductors, demonstrating how the quantum numbers of light-specifically spin angular momentum, orbital angular momentum, and radial order—can be transferred to the superconducting order parameter and control vortex dynamics. Using Laguerre-Gaussian beams, we show that by tuning the quantum numbers and the amplitude of the electric field, it is possible to manipulate a variety of vortex behaviors, including breathing vortex pairs, braiding vortex pairs, vortex droplets
, and swirling 2D vortex rings. More complex structure of vortex-clusters, such as vortex-flake structures, and standing wave motions, also emerge under specific quantum numbers. These results demonstrate the ability to control SC vortex motion and phase structures through structured light, offering potential applications in quantum fluids and optical control of superconducting states. Our findings present a diagram that links light's quantum numbers to the resulting SC vortex behaviors, highlighting the capacity of light to transfer its symmetry onto superconducting condensates. We point that this approach represents the extension of the printing to quantum printing by light in a coherent state of electrons. 
\end{abstract}

\maketitle


\section{\label{sec:intro}Introduction}

Optical printing has evolved significantly over the past few decades. Beyond everyday applications, advanced methods such as laser printing on different materials including bubbles and plasmonic nanoparticles~\cite{zhao2024laser, zhu2016plasmonic}, optical coloration, decolorization, or erasable printing~\cite{zhang2020bioinspired, zhu2017resonant, bocherer2024decolorization, garai2016photochromic}, volumetric printing~\cite{regehly2020xolography, loterie2020high}, and holographic printing~\cite{zhao2020recent} have been reported.

Structured light tailored with quantum numbers (QNs) enables the direct printing of these QNs onto the  condensed electron liquids, and offering precise control over matter states and inducing specific topological and dynamic responses within the material. In condensed matter physics, recent progress in optical control of materials includes  now control of chiral phonon dynamics by the spin angular momentum (SAM) coupling~\cite{Basini2024Terahertz, davies2024phononic, luo2023large},  orbital angular momentum (OAM)  transfer to magnets~\cite{fujita2017encoding}, induction of skyrmions~\cite{parmee2022optical}, induction of spiral Higgs waves~\cite{mizushima2023imprinting}, creation of polariton vortices~\cite{gnusov2023quantum, dominici2023coupled},  Kapitza engineering in superconducting (SC) devices~\cite{yerzhakov2024induction} and even the induction of chiral mass and charge of electrons~\cite{fang2024structured}. These examples showcase the emerging potential of optical manipulation in SC devices and electron behavior.

In superconductors, the coherent quantum electron liquid can be ``stirred up'' by the dynamical electromagnetic gauge potential, which allows for the transfer of the QN of structured light to SC coherent states. In our previous work, ``Structured Light and Induced Vorticity in Superconductors I: Linearly Polarized Light''~\cite{LG_TDGL_I}, we theoretically demonstrated that linearly polarized light can imprint itself on SC fluid inducing transient vortex-antivortex pairs (VPs) due to oscillating magnetic flux provided by the non-homogeneous Gaussianly structured beam. 

\begin{figure*}[!htbp]
    \centering
    \includegraphics[width=0.95\textwidth]{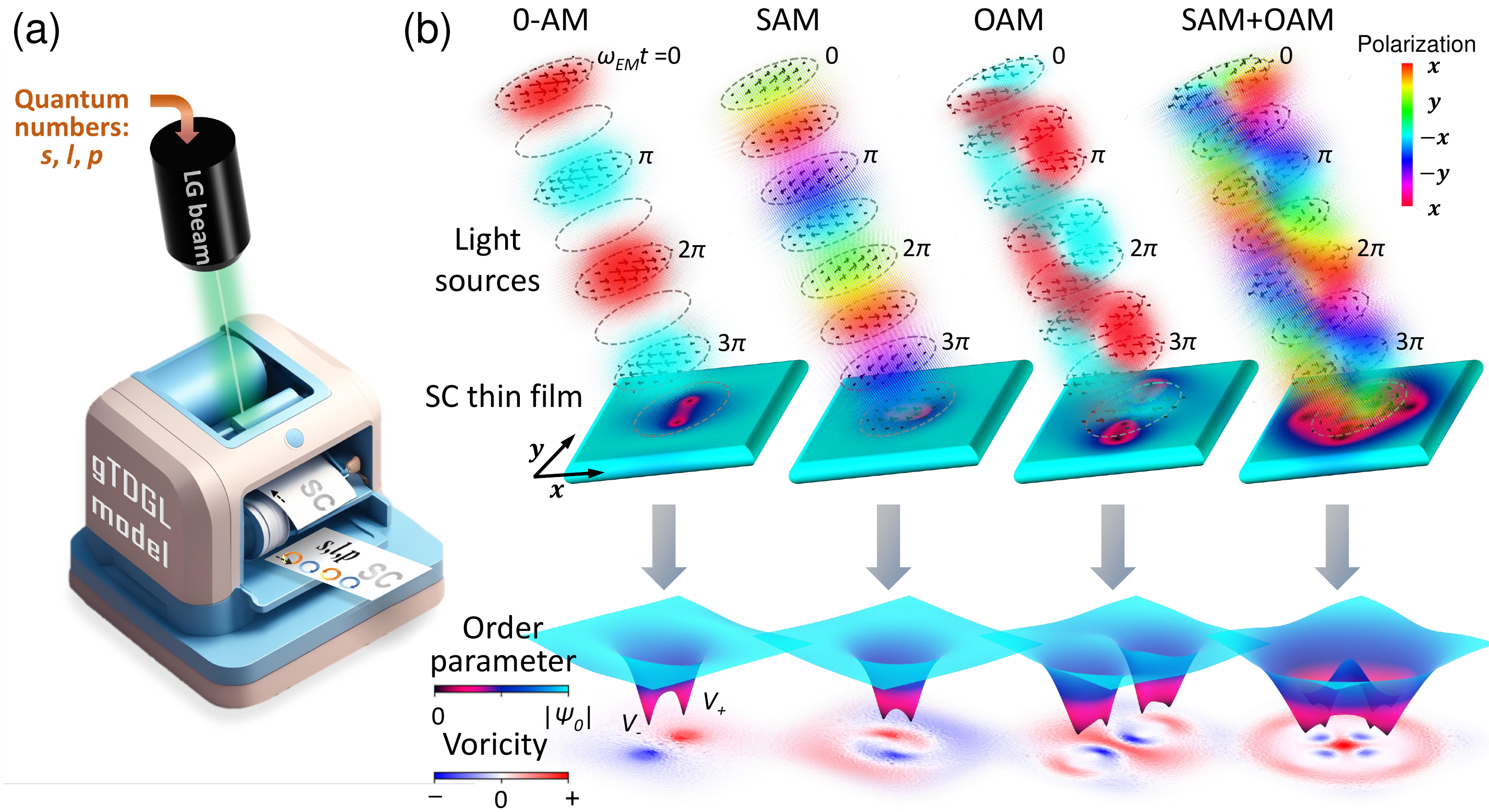}
    \caption{
    (a) A toy model of quantum printing (Partially AI-generated image~\cite{AI}). Within the framework of generalized time-dependent Ginzburg–Landau model (gTDGL), the $s$, $l$, $p$ present the quantum numbers corresponding to the SAM, OAM, and radial order of structured light, respectively. The input and output papers in the printer symbolize the SC thin film before and after imprinting of quantum numbers, respectively. 
    (b) A schematic diagram illustrating the profile of the order parameter and vorticity of the current density induced by structured light. The light propagates normally onto SC thin film. The section of the SC thin film shows the time evolution of light source under different angular momentum (AM) configurations: 0-AM (no AM), SAM ($s=1, l=0$), OAM ($s=0, l=1$), SAM+OAM (combining $s=1$ and $l=1$). The colored region represents the polarization as color bar, with the color intensity. The black dashed circle with arrows inside depicts snapshots of the electric field at different time, $\omega_{EM}t$, where arrows indicate the field direction. The color on the SC surface represents the order parameter as a color map. The figures below the gray arrows show the order parameters affected by structured light and the corresponding vorticity of the supercurrent.}
    \label{fig:abstract}
\end{figure*}

In this work, we extend the investigation to SC coherent states interacting with light carrying non-zero QNs, including SAM, OAM, number of rings of electric field, and mixed effects, as shown schematically in~\cref{fig:abstract}. These QN-carrying light sources are realized through Laguerre-Gaussian (LG) beams. We demonstrate how  structured light carrying SAM and OAM is inducing 
rotating/revolving or braiding vortices. Unlike circularly polarized light, OAM introduces additional out-of-plane flux, resulting in unusual  dynamics. Using a hydrodynamic approach to quantum fluid, we analyze these dynamically imprinted patterns, investigating the supercurrent and vortex responses to QN-carrying light. Our focus is on light sources within the 0.1 to 10 THz range, where wavelengths are on the order of microns or larger. In this series of studies, we simulate the time-dependent motion of SC thin films subjected to incident QN-carrying light using the generalized time-dependent Ginzburg-Landau (gTDGL) model.

We view  the  presented approach of imprinting QN of light onto the coherent electron as an example of {\em quantum print} where coherent fluid of electrons ``senses'' and responds to structured light in a systematic and coherent manner reflecting the possibility to manipulate coherent states with light.

This paper is organized as follows: We show the simulation method in~\cref{sec:method-1}, the light sources in ~\cref{sec:method-2}, and the simulation conditions in~\cref{sec:method-3}. 
We exhibit the resulting dynamics of SC states subjected to light with different SAM and OAM in~\cref{sec:sl}, the influence of QNs which control the number of radial nodes of light, or so-called radial order (RO) or radial index in~\cref{sec:p}. In~\cref{sec:QNs}, we conclude the types of vortex dynamics induced by different QNs.


\section{\label{sec:method}Method: Time-dependent Ginzburg–Landau theory} 

\subsection{\label{sec:method-1}Generalized time-dependent Ginzburg–Landau theory}

We use the generalized time-dependent Ginzburg-Landau (gTDGL) model to explore supercurrent and vortex dynamics. The gTDGL model is described by the following equation in dimensionless units~\cite{kramer1978theory, watts1981nonequilibrium, bishop2023pytdgl},
\begin{multline} 
\label{eq:TDGL}
\frac{u}{\sqrt{1+\gamma^2\left| \psi \right|^2}}
\left(\frac{\partial }{\partial t}+i\mu+\frac{\gamma^2}{2}\frac{\partial \left| \psi \right|^2}{\partial t}\right)\psi
\\ =(1-\left| \psi \right|^2)\psi+(\nabla -i\boldsymbol{A})^2\psi .
\end{multline}
Here, $\psi$ is the complex order parameter, i.e. $\psi=|\psi|\exp{i \theta_{s}}$. 
The term $(\frac{\partial }{\partial t}+i\mu)$ is the covariant time derivative with scalar potential $\mu$, and the term $(\nabla -i\boldsymbol{A})^2$ represents the covariant Laplacian with vector potential $\boldsymbol{A}$ from light sources which will be discussed in the following section. 
The parameter $\gamma$ and the constant $u$ are determined by the influence of inelastic scattering process and the ratio of the relaxation times of the order parameter and current, respectively. 
Further details regarding the equation including used dimensionless units are provided in Ref.~\cite{LG_TDGL_I}.
The total current density is given by:
\begin{equation} 
\label{eq:J}
\boldsymbol{J}=\boldsymbol{J_{s}}+\boldsymbol{J_{n}}=\text{Im}[\psi^{*} (\nabla -i\boldsymbol{A})\psi] - \nabla\mu - \frac{\partial \boldsymbol{A}}{\partial t},
\end{equation}
where $J_s$ and $J_n$ denote the supercurrent density and normal current density, respectively.

The simulation is built upon a Python-based script, LGTDGL~\cite{LGTDGL_github}. Further details regarding the symbols and the gTDGL simulation can be found in Refs.~\cite{LGTDGL_github, LG_TDGL_I, bishop2023pytdgl}.


\subsection{Light sources: Laguerre-Gaussian (LG) beams}
\label{sec:method-2}

Structured light, such as Laguerre–Gaussian (LG) beams, Hermite–Gaussian (HG) beams, and Bessel beams~\cite{forbes2021structured}, offers flexibility not only in intensity profiles but also in their 3D helical phase structures. Typically, linearly polarized light exhibits oscillations in the amplitude of the electric field while maintaining a transverse phase of polarization along one axis. The transverse plane lies in the $x,y$ plane with the propagation direction along the $z$-axis. In contrast, light carrying SAM, such as circularly polarized light, features a rotating transverse phase of polarization and a non-vanishing electric field amplitude. Light carrying OAM exhibits a time-dependent 3D helical phase structure, also with a non-vanishing electric field amplitude except for the place with singularities of transverse phase.
In this work, we use a particular type of OAM carrying beam - a LG beam.
Each LG transverse mode is described by the Laguerre polynomial $L_p^l(f(r))$, where $p$ is the RO, and $l$ is the azimuthal index representing the OAM quantum number. The expression for the amplitude of the electric field of LG mode is given by ~\cite{wang2022orbital, gu2018gouy}: 
\begin{multline} 
\label{eq:LG_u}
u_{l,p}\left(r,\varphi,z\right)= E_0 C_{pl} \left[\frac{\sqrt2r}{w\left(z\right)}\right]^l \frac{w_0}{w\left(z\right)}
\\ L_p^l \left(\frac{2r^2}{w^2\left(z\right)}\right) \cdot e^{-i\phi_{pl}\left(z\right)+i\frac{k r^2}{2q\left(z\right)}+il\varphi} ,
\end{multline}
Here, $p$ determines the number of rings related to spatial intensity of light, and $l$ controls the number of twists of the phase front. The constant $C_{pl}=\sqrt{\mathstrut 2 p!/\pi (p+|l|)!}$ is used to normalize the amplitude.

The $z$-dependent radius of spot size is equivalent to $w(z)=w_0 \sqrt{\mathstrut (z^2+z_R^2)/z_R^2}$ with beam waist $w_0$ and Rayleigh distance $z_R=k \omega_0^2/2$. 
The imaginary terms inside the exponential function include the first term with respect to Gouy phase $\phi_{pl}(z)=(2p+\abs{l}+1) \arctan{(z/z_R)}$, the second term $\exp{ikr^2/2q}$ related to the radius of curvature with $q(z)=z-iz_R$, and the third term $\exp{il\varphi}$ providing spiral wavefront via controlling temporal phase delay $l\varphi$ around the azimuthal angle $\varphi$.
The wavenumber $k$ is $2 \pi/ \lambda_{EM}$ where the wavelength is $\lambda_{EM}=c \tau_{EM}$ and the period of light is $\tau_{EM}$. 
On the focal plane, i.e. $z=0$, the radius of spot size is $w(z)=w_0$, the Gouy phase is 0, and the term of curvature $\exp{i kr^2/2q(z)}$ becomes Gaussian distribution $\exp{-r^2/w_0^2}$.
Throughout this paper, we consider that the sample is placed at focal plane, and the angular frequency of light $\omega_{EM}$ (hereinafter referred to as frequency) is $2\pi/\tau_{EM}$.

Considering the case with polarization along $x$- and $y$-direction and frequency $\omega_{EM}$, the electric field of light carrying SAM number $s$, and the previously mentioned $l$ and $p$, is formulated as:
\begin{multline} 
\label{eq:LG_E}
\boldsymbol{E_{EM}}(\textbf{r}, t) = 
\\ u_{l,p}(r,\varphi,z)\cdot e^{-i(\omega_{EM} t-\varphi_{t0})}
\begin{pmatrix}
\cos{\varphi_{xy}}  \\
 e^{-i \sigma} \sin{\varphi_{xy}} 
\end{pmatrix}
\begin{pmatrix}
{\hat{\mathbf{e}}}_{x} \\
{\hat{\mathbf{e}}}_{y}
\end{pmatrix} ,
\end{multline}
where $\varphi_{t0}$ is the temporal phase at $t=0$, $\varphi_{xy}$ is a constant of azimuthal angle that offsets the polarization along $xy$-plane, and $\sigma=s \cdot \pi/2$. For linear polarization (LP), $s$ is an even number, while for circular polarization (CP),  $s$ is odd. Specifically, we use $s=0$ for LP, $s=1$ for left-handed CP, and $s=-1$ for right-handed CP. We use the notation LG$_{pl,s}$ to denote the mode of LG beam. The parameters of LG beam throughout this work are listed in ~\cref{tab:LGmode}.

\begin{table*}[!htbp]
\centering
\caption{The setting of LG beams mentioned in~\cref{sec:sl} and in~\cref{sec:p}}
\begin{tabular}{lllllll}
\hline
Light source 
& $s$ 
& $l$ 
& $p$
& $\phi_{t0}$
& $\phi_{xy}$
& Description \\
\hline
LG$_{00,0}$ & 0 & 0 & 0 & 0 & 0 & LP Gaussian beam without SAM and OAM ($x$-polarization in this work) \\
LG$_{00,\pm 1}$ & $\pm$1 & 0 & 0 & 0 & $\pi/4$ & CP Gaussian beam with only SAM ($+/-$ for left-/right-handed circular polarization) \\
LG$_{01,0}$ & 0 & 1 & 0 & 0 & 0 & LP LG beam with only OAM \\
LG$_{01,\pm 1}$ & $\pm$1 & 1 & 0 & 0 & $\pi/4$ & CP LG beam with SAM and OAM \\
LG$_{0l,0}$ & 0 & $l>1$ & 0 & 0 & 0 & LP LG beam with higher order of OAM \\
LG$_{0l,\pm 1}$ & $\pm$1 & $l>1$ & 0 & 0 & $\pi/4$ & CP LG beam with higher order of OAM \\
LG$_{pl,0}$ & 0 & $l$ & $p>0$ & 0 & 0 & LP LG beam with high-radial-order \\
LG$_{pl,\pm 1}$ & $\pm$1 & $l$ & $p>0$ & 0 & $\pi/4$ & CP LG beam with high-radial-order \\
\hline
\end{tabular}
\label{tab:LGmode}
\end{table*}

The vector potential $A_{EM}$ of free space structured light is given by $E_{EM}/i\omega_{EM}$. The time-dependent out-of-plane magnetic field,  $B_z(r,t)$, is defined by the curl of vector potential:
\begin{align} 
\label{eq:B_z}
    B_z(\boldsymbol{r},t) = (\nabla\times\boldsymbol{A_{EM}})_z .
\end{align}
For light beams without OAM, such as the LP Gaussian beam LG$_{00,0}$ and CP Gaussian beam LG$_{00,\pm 1}$, the structured distribution induces a time-dependent $B_z(\boldsymbol{r},t)$ which is proportional to the curl of amplitude distribution of $\boldsymbol{E}$ (see~\cite{LG_TDGL_I}). 
For light beams carrying OAM, the azimuthal phase factor $\exp{il\varphi}$ introduces additional contributions to $B_z(\boldsymbol{r},t)$ (see discussion in ~\cref{app:Bz}). Snapshots of electric field and resulting $B_z(\boldsymbol{r},t)$ for different combinations of SAM and OAM are shown in Figs.~\ref{fig:Bz_snapshot1}-\ref{fig:Bz_snapshot3}. Behavior of $B_z$ follows the  distinct pattern for $s=-1,0,1$:

i) $s-l=0$. Both QNs $s$ and $l$ contribute to the rotational properties of light. According to~\cref{eq:LG_u} and ~\cref{eq:LG_E}, the signs of the terms $il \varphi$ and $-i \sigma$ in the phase determine the opposite direction of rotation. When $s=l$  $B_z$ oscillates. For $s=l=0$, it shows a strong linear movement (2-fold symmetry), while for $s=l=1$, $B_z$ exhibits cylindrical symmetry.

ii) $s=0$, $l>1$. In this case, the light carries OAM, but remains LP at each moment. The electric field forms a $2l$-fold rotational flower pattern with radial symmetry, while the $B_z$ field is more complicate, resembling a vertical petal-shaped array. Based on numerical results from ~\cref{fig:Bz_snapshot1} in~\cref{app:Bz}, the dynamics of $B_z$ can be simplified as a spinning petal-shaped array. 

iii)  $s \neq 0$, $l \neq s$. In this case, the light carries both SAM and OAM, resulting in a more complex electric field. The electric field deforms between radial polarization and spiral polarization alternatively. For high-order OAM (e.g. $l>1$ or $l<-1$), the radial and spiral polarizations interweave, forming a pattern with ($\abs{l-s}$)-fold symmetry. As a result, the $B_z$ field behaves like a spinning plate with a flower-like shape, also exhibiting ($\abs{l-s}$)-fold symmetry.

Furthermore, the type of $B_z$ indicates how superconducting states are stirred. In the context of vortex generation, the characteristics of $B_z$ help to elucidate vortex dynamics. 
The numerical results for the $B_z$ profiles are presented in~\cref{app:Bz}.


\subsection{Simulation conditions}
\label{sec:method-3}

In our simulation, we focus on SC thin film with thickness $d \ll \xi$, where $\xi$ is the SC coherence length. As a result, we can disregard variations in the order parameter, current, and vector potential (such as absorptance and penetration depth) along the $z$-direction. Additionally, we neglect screening effects by assuming the London penetration depth or Pearl length is larger than the coherence length, and at least longer than one of sample size and the light spot size.

In type-II SCs, the coherence length $\xi$ is typically shorter than the Pearl length $\Lambda$~\cite{pearl1964current} or the London penetration depth $\lambda_L$~\cite{kittel2018introduction}. Based on this, we adopt niobium (Nb) thin films as a prototype for simulations. According to Refs.~\cite{lemberger2007penetration,draskovic2013measuring}, a 2 nm Nb thin film has $\xi$ in the range of 10$-$100 nm and $\Lambda$ in the range of 1$-$500 $\mu$m. This allows us to simulate EM wave at frequency up to tens of THz, as the beam spot size of the beam has to be on the order of the wavelength.\cite{LG_TDGL_I}
Unless specified otherwise, the simulations assume SC thin films with $L=25-30 \xi$ and $d=0.02 \xi$, which for $\xi=100$ nm~\cite{lemberger2007penetration, draskovic2013measuring} corresponds to $L=2.5-3\mu$m and $d=2$nm. The size of sample is deliberately reduced to enhance computational efficiency. This is specific to the simulation setup and not indicative of real-world application constraints.

Additionally, the sample size $L$ is chosen to be sufficiently larger to avoid edge effect~\cite{LG_TDGL_I}, particularly for light beams with higher RO. The beam spot size $2 w_0$ is fixed at $12 \xi$. The intensity of light varies with each quantum number, which will be discussed later. In~\cref{app:E}, we provide the values of $E_0$ used in the simulations and the corresponding sample size $L$, except for the the specific cases which are mentioned in the figure captions. The mesh of simulation is smaller than $\xi$, which is discussed in~\cref{app:E}. Besides, we set $\gamma=10$ and $\epsilon=0$ for the simulation.


\section{\label{sec:results}Results}

\subsection{Spin and orbital angular momentum}
\label{sec:sl}

Light carrying SAM and OAM imparts rotational and revolving orbital-motion to the superconducting coherent state. 
Vortex generation through the AM transfer from an electromagnetic beam during processes of hot spot quenching and superconducting condensate formation has been explored in Refs.~\cite{yokoyama2020creation, plastovets2022all}. In contrast, our previous work~\cite{LG_TDGL_I} demonstrates that vortices can also be generated without the need for AM transfer, and independent of the quenching process.
The resulting out-of-plane $B_z$ induces and stirs the supercurrent, affecting both the fast response of $\theta_s$ and slower reaction of $\abs{\psi}$ ~\cite{LG_TDGL_I}. In this section, we begin by examining five simple cases of SAM and OAM, ($s$,$l$)=(0,0), (1,0), (0,1), (-1,1), and (1,1), to explore the transfer of AM from light to vortices.
Following our previous work\cite{LG_TDGL_I}, we use the light frequency at $\omega_{EM}=\omega_{GL}/40$ to allow sufficient time for the generation and recombination of VPs.

\subsubsection{Imprint effect of supercurrent}
\label{sec:imprint}

In the standard TDGL model, the unit time of system $\tau_{GL}$ is determined by the relaxation time of the suprcurrent~\cite{kopnin2001theory}, while the relaxation time of order parameter is $u$ times longer than $\tau_{GL}$. In the gTDGL model, the relaxation time of the order parameter is even longer. When considering the relaxation times of $\abs{\psi}$ and $\theta_s$ induced by the optical pulse, they can be $\sim 30 \tau_{GL}$ and $\sim 1.5 \tau_{GL}$ for $\gamma=10$, respectively (see Appendix B in previous work~\cite{LG_TDGL_I}). 
It clearly demonstrates the rapid response of the supercurrent and $\theta_s$, followed by the delayed response of $\abs{\psi}$. The structured light imprints its effects first on the supercurrent directly, and induces the striped structure on $\theta_s$ before vortex generation, with the longest-lasting influence observed in $\abs{\psi}$.

\begin{figure*}[!htbp]
    \centering
    \includegraphics[width=0.85\textwidth]{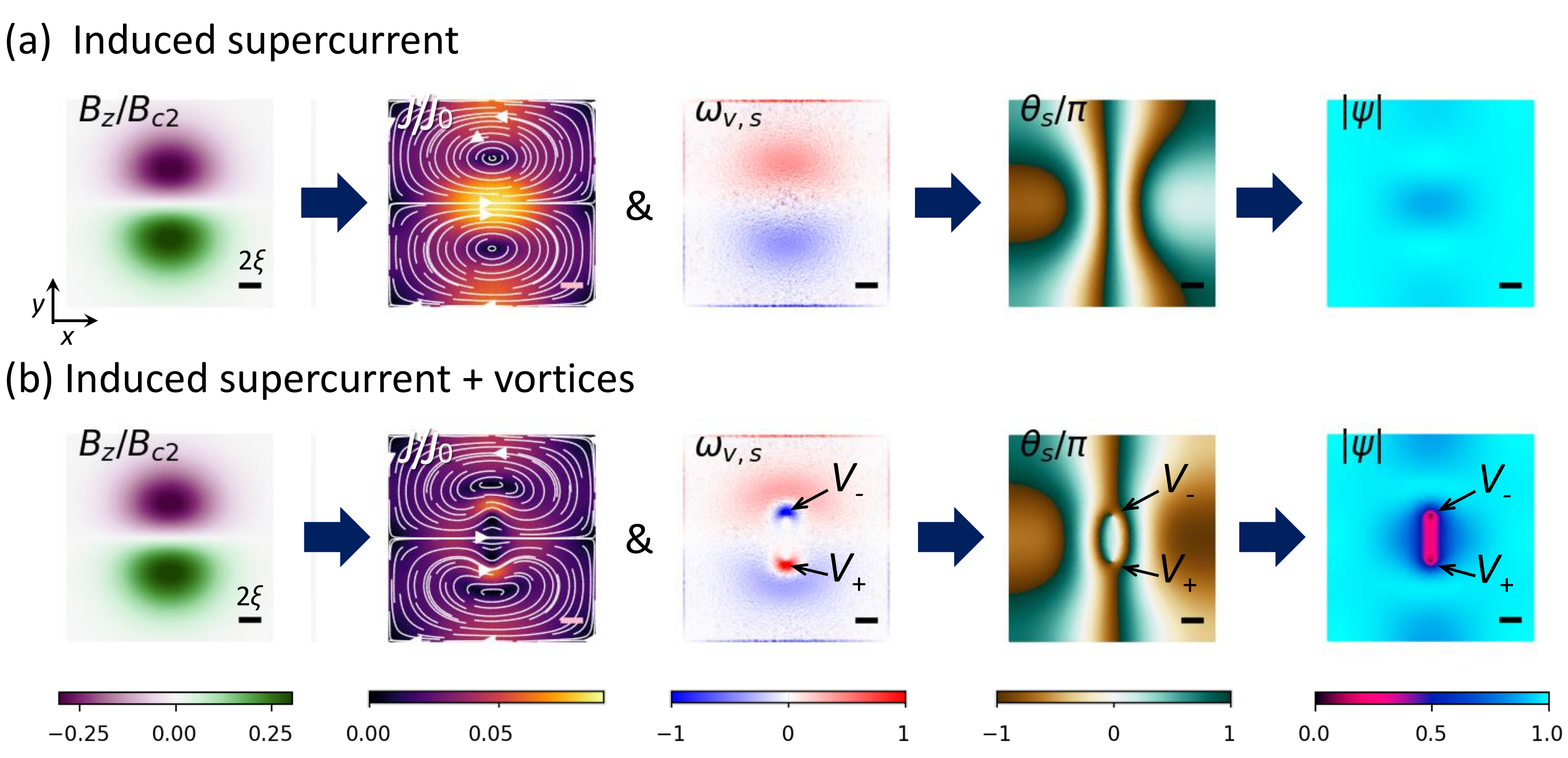}
    \caption{Different profiles showing the effects of optically induced supercurrents and vortices. These profiles include out-of-plane magnetic field $B_z$, supercurrent density $J_0$, vorticity of supercurrent density $\omega_{nu,s}$, phase of order parameter $\theta_s$, and amplitude of order parameter $\abs{\psi}$, from left to right. The arrows indicate the transfer order from $B_z$ of light through $J_0$, normalized $\omega_{nu,s}$, $\theta_s$, to $\abs{\psi}$. Snapshot of (a) shows the LG$_{00,0}$ induced supercurrent before vortices generation at $t=10\tau_{GL}$, while (b) shows the state after vortices generation at $t=90\tau_{GL}$. The notations $V_+$ and $V_-$ represent vortex and antivortex, respectively.}
    \label{fig:supercurrent}
\end{figure*}

If we define the curl operator as representing the vorticity of a physical parameter, then $B_z$ can be interpreted as the vorticity of the vector potential, as expressed in~\cref{eq:B_z}.
~\cref{fig:Bz_Wn1_A1,fig:Bz_Wn1_A2,fig:Bz_Wn1_A3} in~\cref{app:vorticity} provide the numerical evidence that the vorticity profile of the supercurrent density $\omega_{\nu,s} = \nabla\times\boldsymbol{J_{s}}$ directly inherits its structure from the $B_z$ field, occurring before vortex generation. 
Assuming that the contributions from the phase of order parameter $\theta_s$, including pure supercurrent component $\theta_{s,Js}$ and vortex component $\theta_{s,V}$, can be separated, i.e., $\theta_s \approx \theta_{s,Js} + \theta_{s,V}$, and neglecting the contribution from the variation in the amplitude of the order parameter, the total vorticity of the supercurrent $\boldsymbol{J_s}=\text{Im}[\psi^{*} (\nabla -i\boldsymbol{A})\psi]=\abs{\psi}^2 (\nabla\theta_s -\boldsymbol{A})$ (see~\cref{eq:J}) 
can be approximated as
\begin{multline} 
\label{eq:curl_Js}
    \nabla\times\boldsymbol{J_s}
    \approx\abs{\psi}^2 \nabla\times(\nabla\theta_{s} -\boldsymbol{A})
    = \abs{\psi}^2 \left( \nabla\times\nabla\theta_{s} -\nabla\times\boldsymbol{A} \right) \\
    \approx \abs{\psi}^2 \left( \nabla\times\nabla\theta_{s,Js} + \nabla\times\nabla\theta_{s,V} -\nabla\times\boldsymbol{A} \right).
\end{multline}
Since the phase $\theta_{s,Js}$ provides the continuous part of the phase, the first term in the brackets in the last line of~\cref{eq:curl_Js} turns out to be $\nabla\times\nabla\theta_{s,Js}=0$. The second term is a non-zero value due to the singularity point of $\theta_{s,V}$ in vortex. The third term is equivalent to $-B_z$ according to~\cref{eq:B_z}. Hence, before the vortex generation, the vorticity of supercurrent can be rewritten as 
\begin{equation} 
\label{eq:curl_Js_current}
    \omega_{\nu,s}=\nabla\times\boldsymbol{J_s} \approx -\abs{\psi}^2 \nabla\times\boldsymbol{A} = -\abs{\psi}^2 B_z .
\end{equation}
The proof in ~\cref{eq:curl_Js} and~\cref{eq:curl_Js_current} directly demonstrate the imprinting effect, where the vorticity of the vector potential is transferred to the vorticity of the supercurrent, leading to the observed dynamics.

We demonstrate this in~\cref{fig:supercurrent}, where snapshots of the superconducting states and magnetic fields at times before and after generation of a VP are shown for the case of linearly polarized light.
In~\cref{fig:supercurrent} (a), the vorticity $\omega_{\nu,s}$ of the supercurrent is shown as a faint red area for positive vorticity and a faint blue area for negative vorticity. The resemblance between plots for $B_z$ and $\omega_{v,s}$ is clearly visible. This resemblance persists for all types of structured light (~\cref{app:vorticity}). 
After the vortices emerge, the vorticity manifests as bright red or blue dots, representing vortices ($V_+$) and antivortices ($V_-$). The profile of $\theta_s$ shows singularity points with opposite winding numbers, while profile of $\abs{\psi}$ exhibits zeros at the location of the vortices.

In contrast to the profile of the supercurrent, the motion of vortices is more complex and unpredictable. Therefore, in the following discussion, we will focus on the dynamics of vortex motion.


\subsubsection{Imprint effect of spin angular momentum $s$}
\label{sec:sl_SAM}

As was in details shown in the case of linearly polarized light, LG$_{00,0}$, the VPs are generated due to oscillating in time locally non-zero $z$-component of magnetic field that changes sign spatially and creates regions of positive and negative magnetic fluxes. When these fluxes greater than magnetic flux quantum $\Phi_0$, vortex-antivortex pairs might be generated.
Vortex dynamics for several cases of light carrying SAM and OAM is shown in~\cref{fig:vortex_3D}. The light sources LG$_{00,0}$ and LG$_{00,\pm 1}$ serve as the first examples, both exhibiting similar transient $B_z$ profiles but with different dynamical behaviors. The former carries zero AM as a LP beam, while the latter carries SAM $s=1$ for left-handed CP and $s=-1$ for right-handed CP. 
As exhibit in~\cref{fig:Bz_snapshot1} of~\cref{app:Bz}, the LP LG$_{00,0}$ exhibits oscillating dual-semicircle patterns in $B_z$, whereas the CP LG$_{00,\pm 1}$ displays spinning dual-semicircle patterns in $B_z$ with constant field intensity. ~\cref{fig:vortex_3D} illustrates the results of vortex dynamics in (a) for LG$_{00,0}$ and (b) for LG$_{00,1}$, also presenting snapshots of the order parameter amplitude $\abs{\psi}$, phase $\theta_s$, and vorticity of the current $\omega_{\nu,s}$. The vortex time traces are defined by the regions where $\abs{\psi}<0.05$ with the sign of vorticity distinguishing the vortex (red) and antivortex (blue). 

\begin{figure*}[!htbp]
    \centering
    \includegraphics[width=1\textwidth]{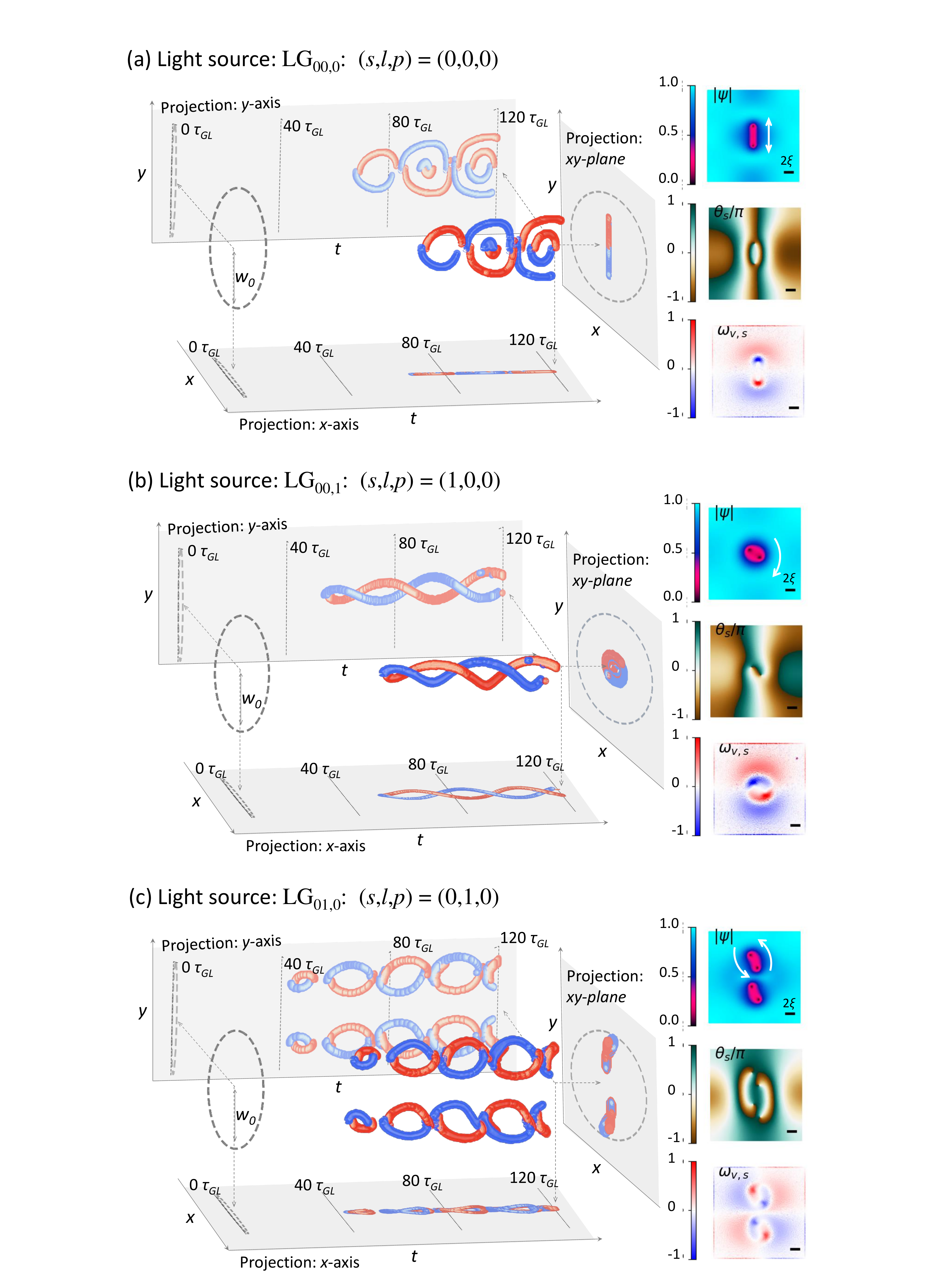}\clearpage
    \caption{(Full caption on the next second page.)}
\end{figure*}

\begin{figure*}[!htbp]
    \centering
    \ContinuedFloat
    \includegraphics[width=1\textwidth]{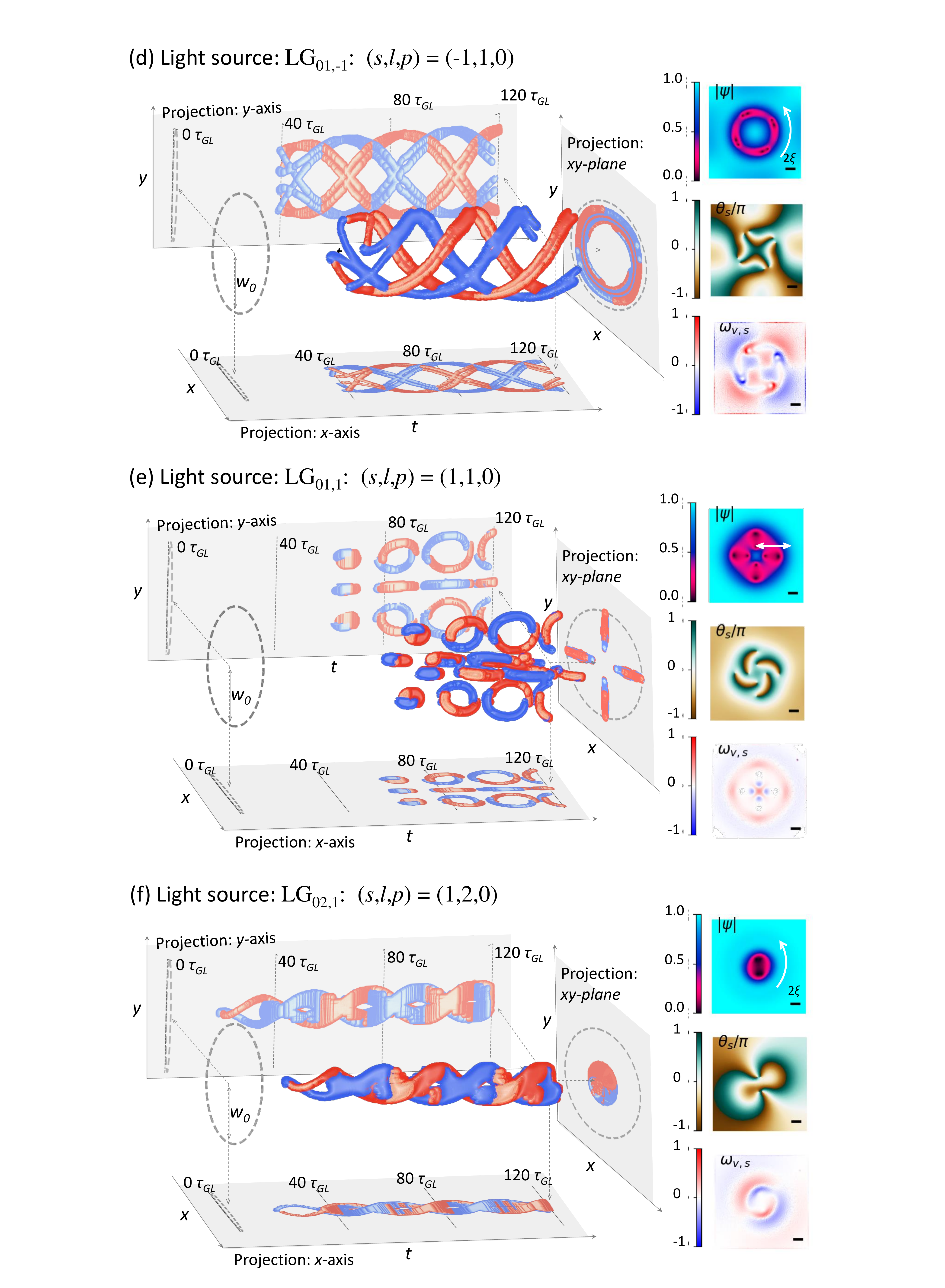}\clearpage
    \caption{(Full caption on the next page.)}
\end{figure*}

\begin{figure*}[!htbp]
    \centering
    \ContinuedFloat
    \includegraphics[width=1\textwidth]{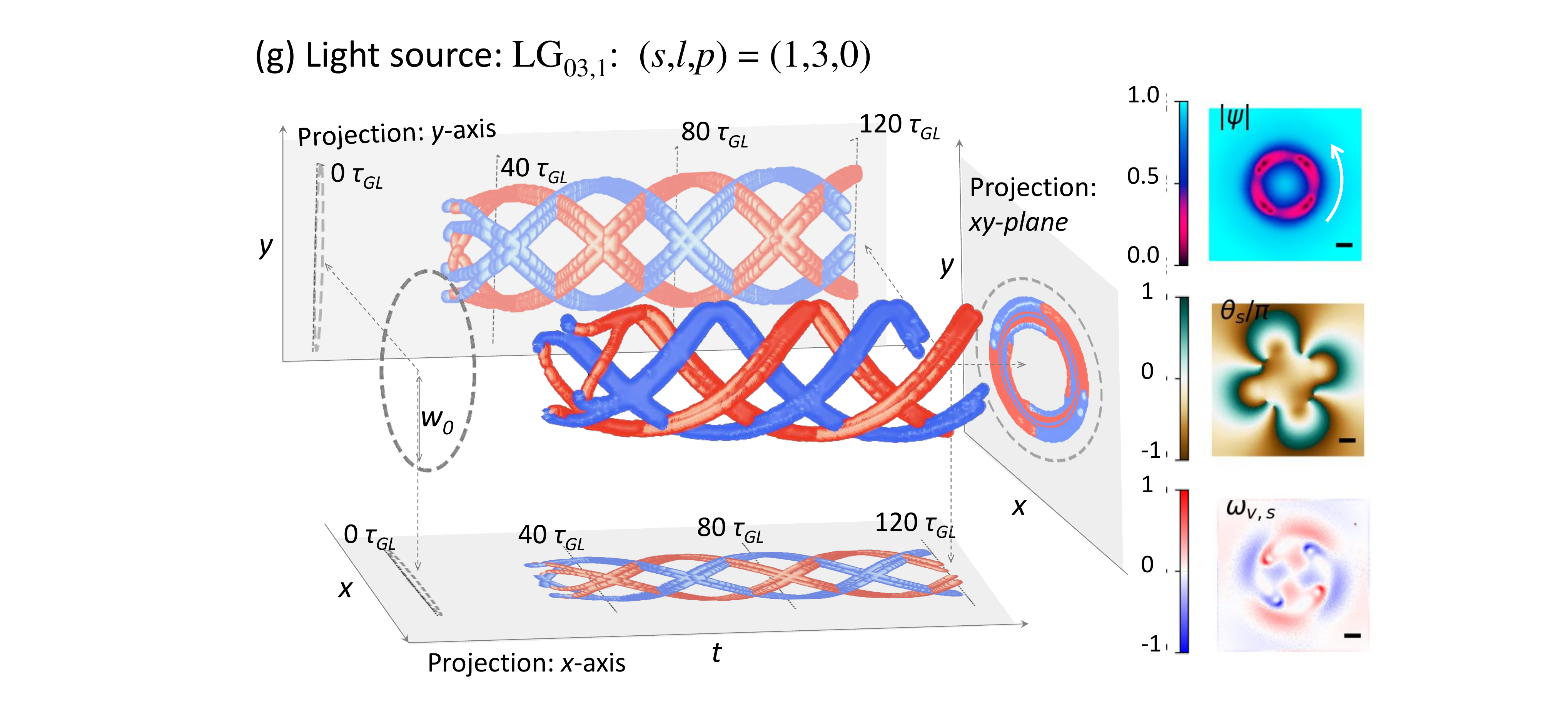}
    \caption{Time-trace of vortices and snapshot of order parameters induced by light sources: (a) LP LG$_{00,0}$, (b) CP LG$_{00,1}$, (c) LP LG$_{01,0}$, (d) CP LG$_{01,-1}$, (e) CP LG$_{01,1}$, (f) CP LG$_{02,1}$, (g) CP LG$_{03,1}$. We illustrate the time-varying 2D traces of vortices. Each time-frame of 2D vortex traces are displayed along vertical slices, like the orientation of the ``projection of $xy$-plane'', while the horizontal direction corresponds to the evolution in time. 
    The red and blue curves depict the trace of vortices and antivortices, respectively. The inserts of $\abs{\psi}$, $\theta_s$, and $\omega_{\nu,s}$ are the snapshot of amplitude of order parameter, phase, and vorticity of the supercurrent, respectively. The full motion records are available in the Supplementary materials~\cite{supp1, supp2, supp3}.}
    \label{fig:vortex_3D}
\end{figure*}

The vortex motion exhibits distinct characteristics depending on the light beam. The LP LG$_{00,0}$ generates staggered vortex and antivortex pairs that appear and disappear periodically, with intervals between the processes of generation and recombination (as seen in ~\cref{fig:vortex_3D}(a)). In contrast, the CP LG$_{00,1}$ produces braided VP (as shown in ~\cref{fig:vortex_3D}(b)) without recombination. 
Due to SAM, the VP induced by LG$_{00,1}$ also spins around their point of origin. The vortex dynamics induced by LP LG$_{00,0}$ have been thoroughly discussed in our previous work~\cite{LG_TDGL_I}. In contrast, for light carrying SAM, unlike the case of zero AM, once the system begins to rotate, the braided VP can move perpetually, presenting a ``yin-yang'' pattern of vorticity comprised of the opposing vorticity of the vortices of the VP within the supercurrent.
The evolution of the vortex pairs is further clarified by their projection on the $xy$-plane, which shows linear scars (or the trace/traces of vortices) for the LP LG$_{00,0}$ beam and circular scars for the CP LG$_{00,1}$ SAM beam. The complete dynamics of the LG$_{00,0}$ and LG$_{00,\pm 1}$ sources are demonstrated in the Supplementary materials~\cite{supp1, supp2, supp3}.


\subsubsection{Imprint effect of orbital angular momentum $l$}
\label{sec:sl_OAM}

OAM, which imparts a revolving motion to light rather than spinning, affects the order parameter and superflow in a distinct way. The rotational phase factor $\exp{i l \phi}$ around the center leads to a null electric field at the beam's center and contributes to oscillating in time $B_z$. In~\cref{fig:vortex_3D}(c), we present the time evolution of vortices and order parameters induced by light carrying pure OAM, $l=1$. The vortex traces and $\abs{\psi}$ display two linear-like motions of VPs in the upper and lower ($y \gtrless 0$) half-planes. 

The mechanism of vortex generation in this context parallels the process observed in vortex generation induced by the linearly polarized light~\cite{LG_TDGL_I}, where vortices and antivortices emerge from regions with zero magnetic field, $B_z=0$, positioned between the maxima and minima of $B_z$. However, instead of generating a branch cut between paired vortices, each vortex in this configuration links its branch cut to the vortex located in a different VP. This arrangement makes the phase profile $C_2$ rotationally symmetrical about $(x=0,y=0)$ point (center of the sample). During the vortex generation, the rotating magnetic field component $B_z$ ``tears up'' the phase from left and right sides, and further displaces the phase between them, like the shear force along $x=0$ line (transform boundary). In~\cref{fig:phase_front_shear}(a) (see~\cref{app:shear}), we demonstrate the phase displacement and subsequent VP formation, while (b) shows the phase recovering as the VPs recombination.


Combining OAM and SAM introduces more complexity. 
In the case of the beam with OAM number $l=1$ paired with the left- or right-handed CP, the resulting patterns of both $B_z$ and vortex dynamics vary significantly.
For $s=-1$, LG$_{01,-1}$, when the SAM aligns with the OAM, $B_z$  exhibits a 2-fold rotational symmetry due to $l-s=2$. ~\cref{fig:vortex_3D}(d) and Supplementary material~\cite{supp2} present the vortex motions and dynamical simulation results.
Vortex generation begins with four VPs around the ring-orbital, continuing in a perpetual revolving motion. Besides, the resulting $\theta_s$ profile shows concave branch cuts.

For the case of $s=1$, LG$_{01,1}$ corresponds to the opposite directions of SAM and OAM. The vortex motion and a snapshot of the SC state are presented in~\cref{fig:vortex_3D}(e), and the vortex motions and full dynamics of the SC condensate are provided in the Supplementary material~\cite{supp3}. The electric field $\b{E}$ exhibits dynamical changes between radial and spiral polarization, whereas oscillating in time $B_z$ is spherically symmetric with a nodal line (circle at which $B_z=0$).
Strikingly, the vortices move linearly, similar to vortex motion induced by linear light (see~\cref{fig:vortex_3D}(a)), but distributed radially. In contrast, the branch cuts in the SC phase profile alternate between radial and spiral alignment, resembling the behavior of $\b{E}$ of the light beam.
The spinning nature appears to be concealed in the $\theta_s$ profile and remains invisible in $\abs{\psi}$.
 
For this case, the resulting linear motion of vortices in the VPs can be explained using the concept of inverse trammels (see details in~\cref{app:shear}).
The traditional trammel method provides a way to plot a circular (or elliptical) trajectory using linear motion. In the vortex generation case for $s=l=1$, when the amplitude $E_0$ is sufficient to generate 4 VPs (the minimum number of VPs LG$_{01,1}$ can generate),  
$B_z$ field induces shift of phases along two intersecting lines, as shown by the gray cross in~\cref{fig:trammels}(a).
The phase rotates along in these lines, driving the linear generation and recombination of VPs. 
The four occurring branch cuts are analogous to four trammels along the intersecting lines, effectively illustrating an inverse of trammel process, where circular motion is converted into linear motion. Additionally, $B_z$ shows a maximum at center of LG$_{01,1}$, unlike LG$_{00,0}$, which peaks at the sides. Notably, no vortex is present at the center for LG$_{01,1}$. 
In~\cref{fig:trammels}(b), we capture the moment of VP generation as the phase branch cuts displace along the trammel trace (transform boundary), and VP recombination occurs as the displaced phases realign during the swinging motion of ``phase trammels''.

Moreover, the supercurrent $J_s$ and vorticity $\omega_{\nu,s}$ in~\cref{fig:vortex_3D}(e) and the Supplementary material~\cite{supp3} exhibit a clearly swirly supercurrent along the orbital which behaves like a closed-loop circuit, but it is purly induced by the light source LG$_{01,1}$. The swirly supercurrent flips the sign immediately as $B_z$ changes the sign. This supercurent behaves like a super-vortex, which can be also found in the cases of LG$_{11,1}$ and LG$_{21,1}$ in Supplementary materials~\cite{supp6, supp9}.

Consequently, the results from LG$_{01,1}$ and LG$_{01,-1}$ suggest that the QNs of SAM and OAM are interconnected in light and cannot be treated independently. While the supercurrent and vortex motion are not directly related to the QNs, they are more easily associated with the behavior of $B_z$. Additionally, the formation of vortices requires a minimum number of vortex pairs, implying that a higher field intensity $E_0$ is necessary for their generation. In the next section, a general case for QNs will be discussed, offering a broader perspective on imprinting.


\subsubsection{Higher order of orbital angular momentum}
\label{sec:l>1_OAM}

\paragraph{Quantum numbers $l>1, s=0$} \mbox{}

~\cref{fig:sl}(a) shows the SC states induced by pure OAM-carrying light sources. From $l=1$ to $l=5$, various vortex traces appear in the $\abs{\psi}$ profile, resembling shapes such as colons, lightning bolts, and round brackets. When $l=1,2$, the transform boundary of displacing phase are still observable. For $l>2$, the vortex traces stabilize into round bracket-like shapes. However, in the SC phase profile, the transform boundary becomes difficult to be identified shear lines associated with these round bracket traces. Despite not forming complete circular patterns, the vortices continue to move counterclockwise, following the rotational direction of the OAM.

\begin{figure*}[!htbp]
    \centering
    \includegraphics[width=0.8\textwidth]{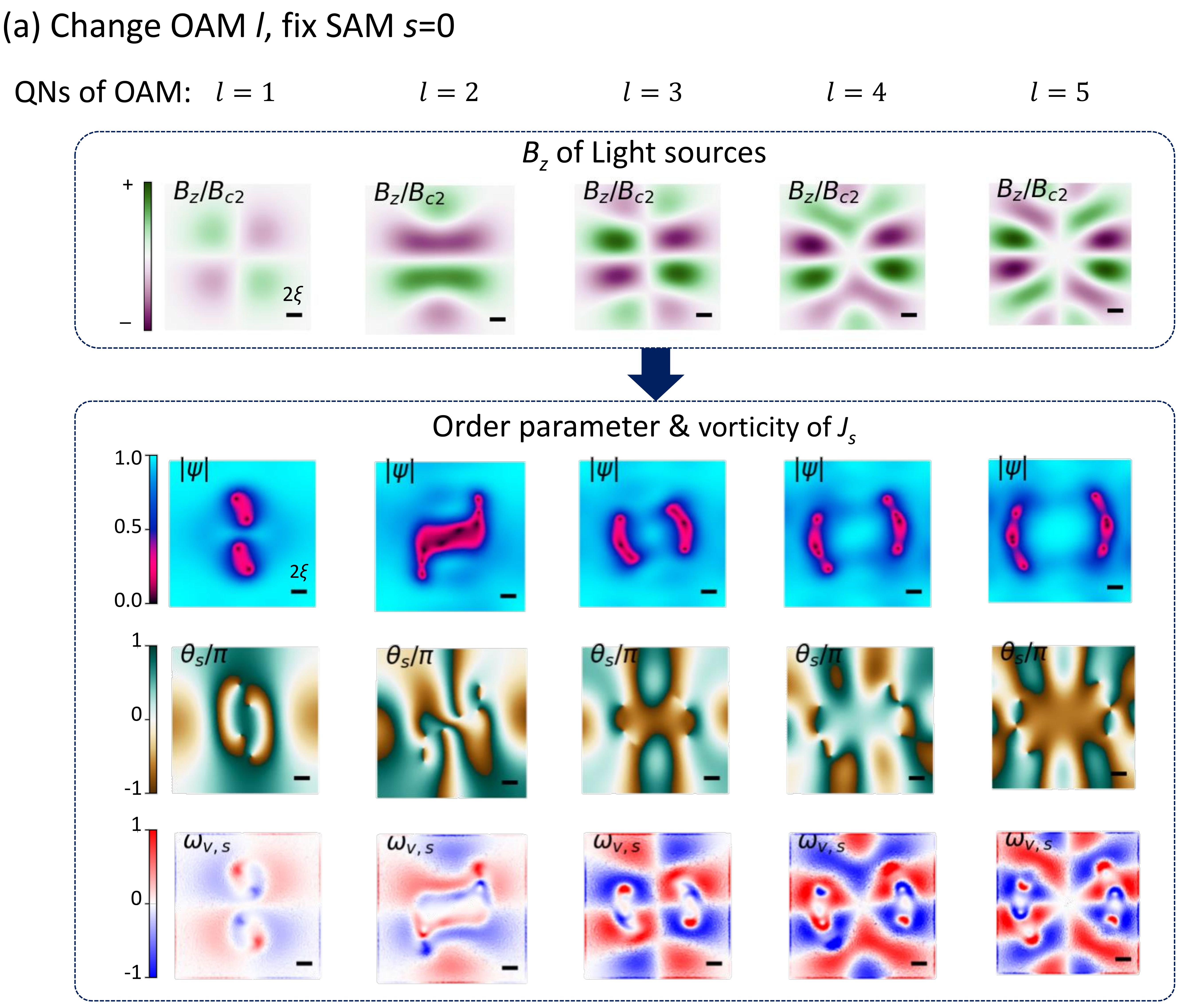}
    \includegraphics[width=0.8\textwidth]{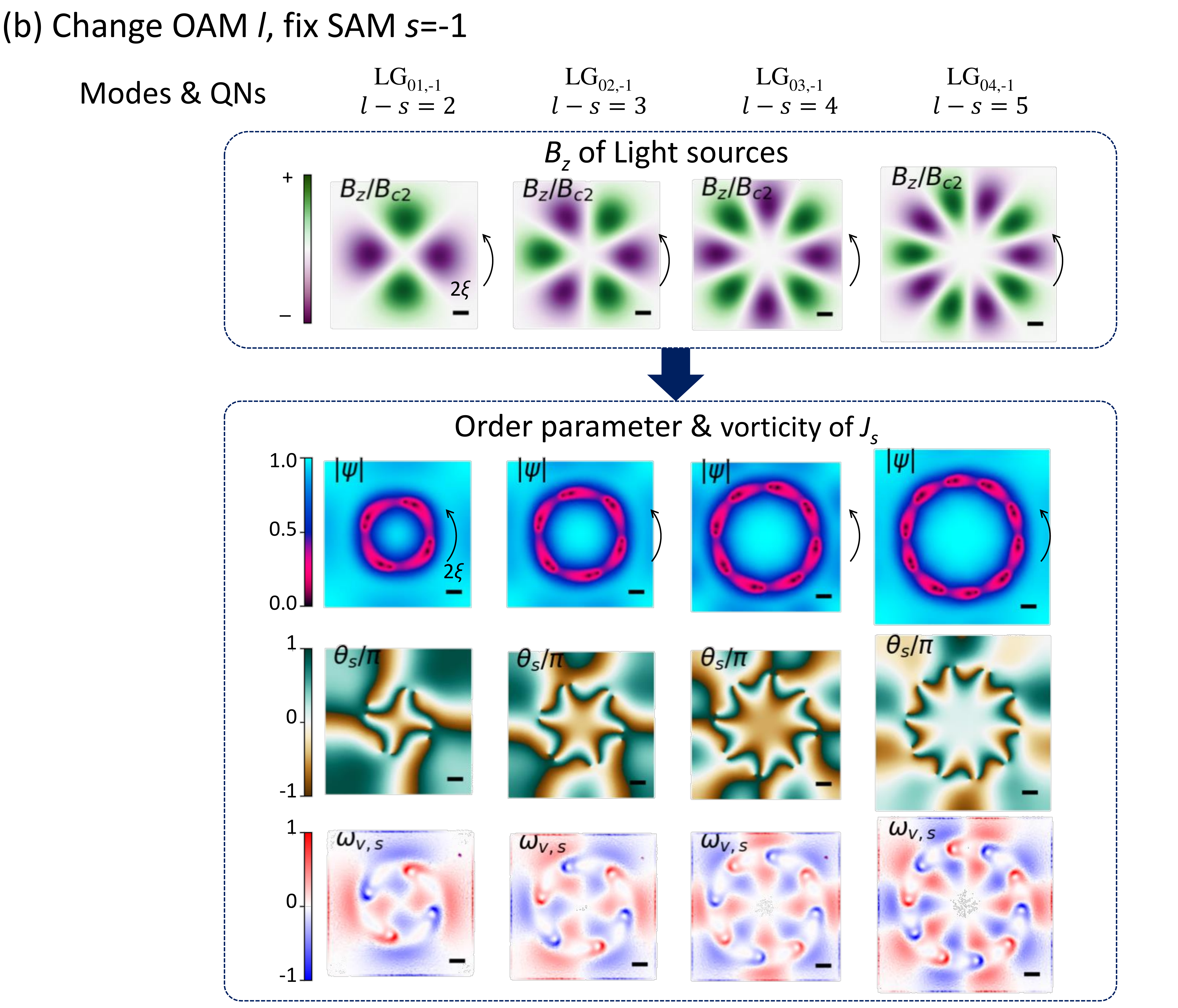}
    \caption{(Full caption on the next page.)}
\end{figure*}

\begin{figure*}[!htbp]
    \centering
    \ContinuedFloat
    \includegraphics[width=0.8\textwidth]{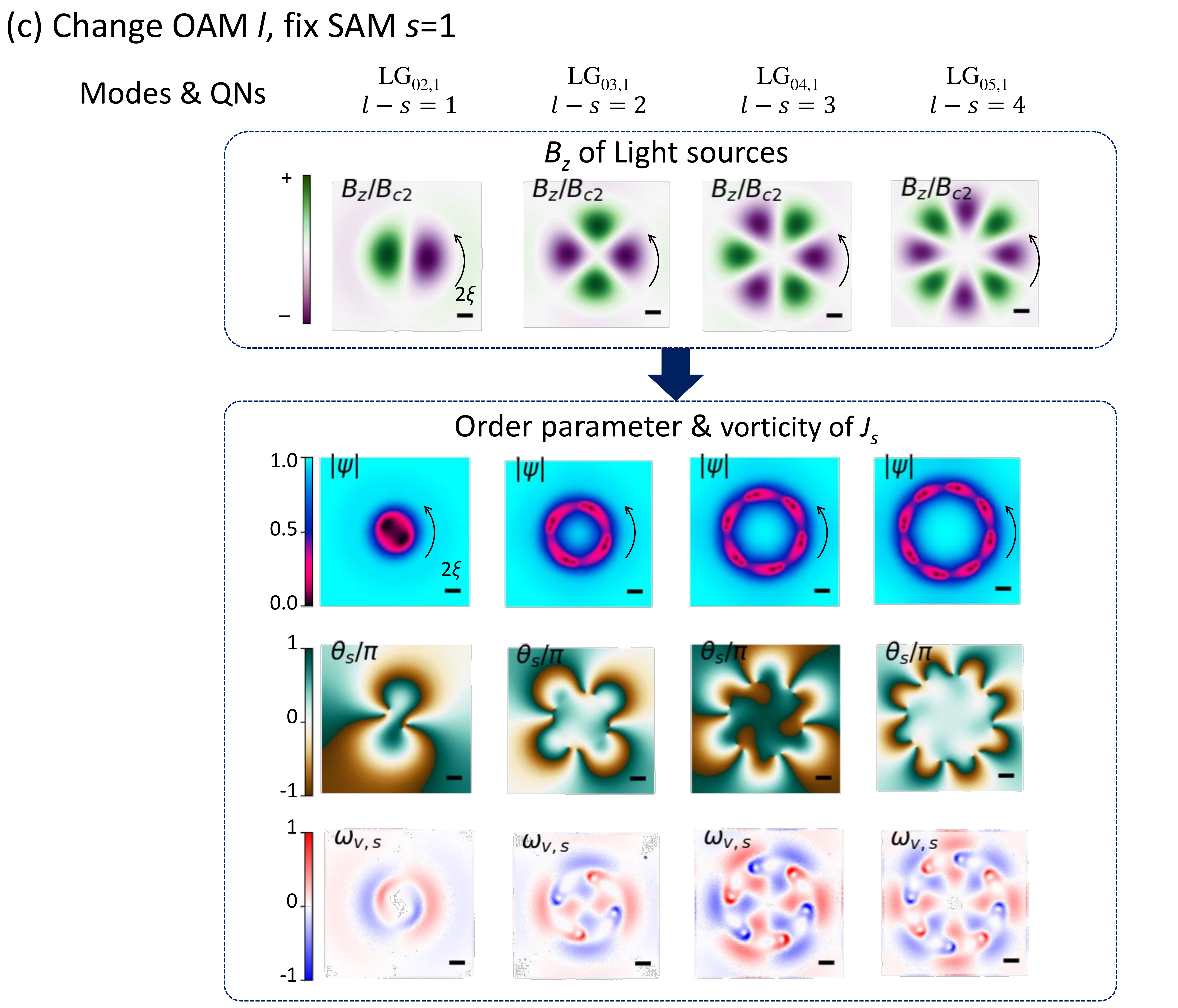}
    \caption{Snapshot of $B_z$ and order parameters induced by higher-order OAM-carrying light sources of quantum numbers (a) from $l=1$ to $l=5$ at $t=90\tau_{GL}$, (b) with left-handed CP at $t=120\tau_{GL}$ and (c) with right-handed CP at $t=120\tau_{GL}$. 
    The imprinted SC states are shown in the lower block, including $\abs{\psi}$, $\theta_s$, and normalized $\omega_{\nu,s}$ from top to bottom as labels. 
    Details of the applied electric fields are provided in~\cref{app:E}. The complete records of the vortex dynamics are available in the Supplementary materials~\cite{supp1, supp2, supp3}.}
    \label{fig:sl}
\end{figure*}

\paragraph{Quantum numbers $l>1, s \neq 0$} \mbox{}

When SAM is present, both $\b{E}$ and $B_z$ exhibit ($\abs{l-s}$)-fold rotational symmetry, with the sign of $(l-s)$ determining the direction of rotation—negative for right-handed and positive for left-handed. 
~\cref{fig:sl}(b) and (c) shows $B_z$ for QNs $l>1, s=-1$ and $l>1, s=1$ along with the corresponding imprinted SC states, respectively. Aside from the ring size effected by $l$, this symmetry is directly imprinted onto the profiles of $\omega_{\nu,s}$ and $\theta_s$. The vortices in $\abs{\psi}$ display the (2$\abs{l-s}$)-fold symmetry because vortices and antivortices are indistinguishable in $\abs{\psi}$.
From a symmetry perspective, both $s=-1$ (left-handed CP) and $s=1$ (right-handed CP) behave similarly when $(l-s)$ are the same.

\paragraph{Convex and concave curves of phase branch cuts} \mbox{}

Analyzing the cases where $(l-s)=2, 3, 4$ in~\cref{fig:sl}(b)(c), we find that the profiles of $\abs{\psi}$ and $\omega_{\nu,s}$ appear nearly identical for the same $(l-s)$ values, with only minor differences in ring size. However, the phase profiles reveal distinctly different features.
Both configurations maintain ($\abs{l-s}$)-fold symmetry with singularity points aligned along the ring orbitals. For $s=-1$, the phase exhibits concave phase branch cuts between each VPs, curving inward toward the inner ring. Conversely, when $s=1$, the phase shows the convex phase branch cuts between the VPs, curving outward toward the outer ring. This difference in phase structure may influence the size of the rings; the convex branch cuts compress the ring, while the concave branch cuts can cause the ring orbital to expand.

Both $s=-1$ and $s=1$ exhibit kaleidoscopic profiles for $\theta_s$, with distinctive features both inside and outside the ring that enrich the overall pattern. The phase branch cuts cause patterns with a high contrast ratio along the ring, in stark contrast to the smooth patterns outside of it.
Within the phase branch cuts, plateau ($\theta_s \nsim 0$) or valley ($\theta_s \sim 0$) structures, those features maintain rotational symmetry in the phase branch cuts. The feature of outer phase rotates as the vortex move while the inner constant structure oscillates in amplitude during the revolution of the VPs. This components complicate the phase structure, yet they are nearly muted in the $\abs{\psi}$ profile. Due to these characteristics, the $\theta_s$ profile for $s=-1$ results in a ($\abs{l-s}$)-fold star-shaped phase, while $s=1$ yields a ($\abs{l-s}$)-fold flower-shaped phase. The complete motion is displayed in Supplementary materials
~\cite{supp2, supp3}.

\paragraph{Vortex rings and vortex droplet rings} \mbox{}

In addition to the phase structure, distinct vortex structures can be observed in the profiles of $\abs{\psi}$. For the cases of LG$_{02,1}$ and LG$_{03,1}$, each node of the ring exhibits two singularity points in the $\theta_s$ profile. However, LG$_{02,1}$ contains one hole in the $\abs{\psi}$ profile (i.e. the area of $\abs{\psi}\sim 0$), while LG$_{03,1}$ has two holes. Based on the morphology of the order parameters, these behaviors can be classified as braiding vortex droplets for LG$_{02,1}$, and swirling 2D vortex ring for LG$_{03,1}$.

The demonstration in~\cref{fig:vortex_3D}(f)(g) illustrates the vortex traces of both vortex-droplets and vortices. Notably, aside from the quantity and size of vortices, the size of the orbital is a key factor influencing this morphology.

\begin{figure*}[!htbp]
    \centering
    \includegraphics[width=1\textwidth]{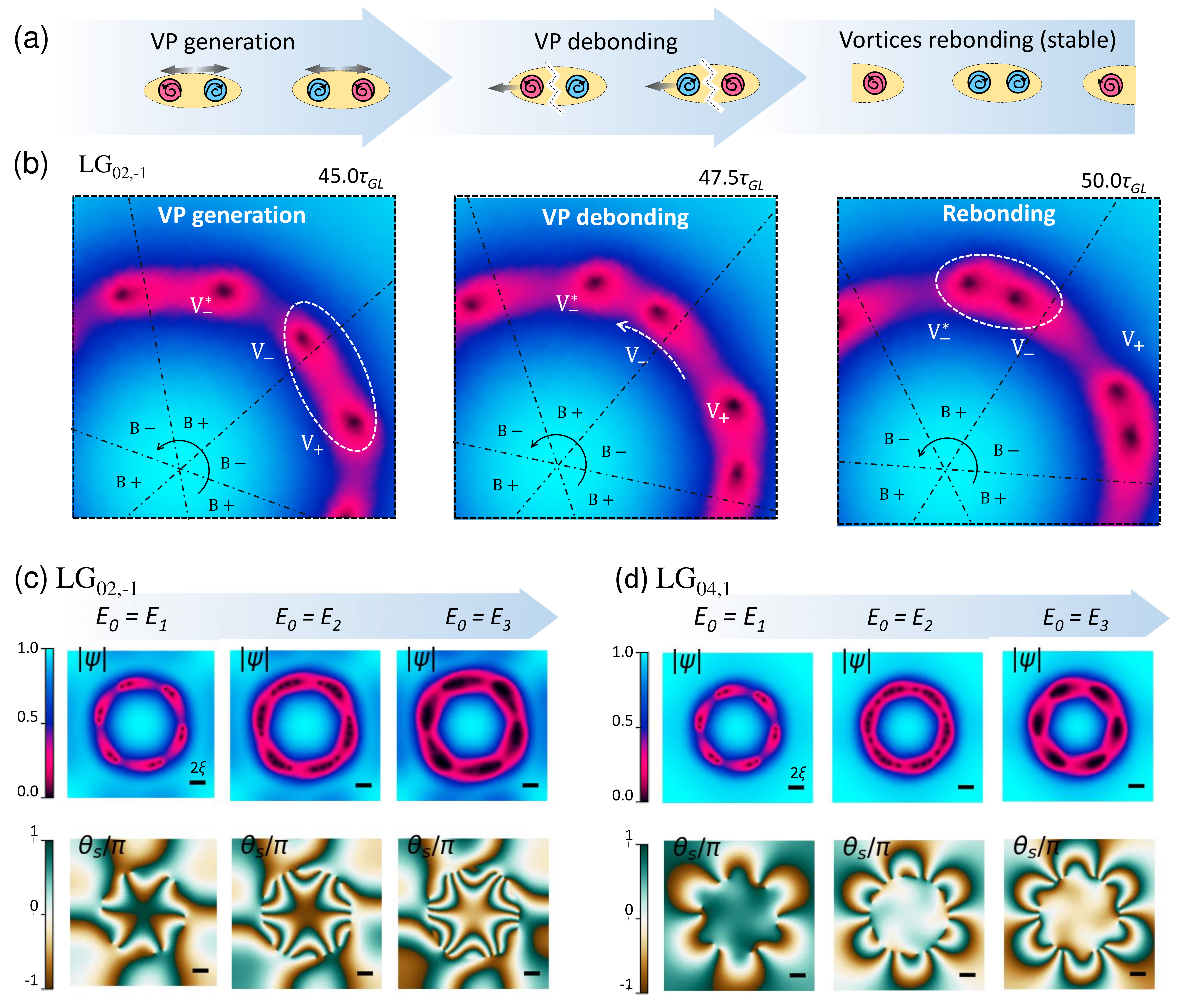}
    \caption{(a)(b) are related to vortex debonding, rebonding process, and (c)(d) are in regard to formation of vortex-droplet ring. (a) Schematic diagram of debonding and rebonding process. (b) Tracking the formation of vortex rings. The light source is LG$_{02,-1}$ for demonstration. Snapshots from top to bottom illustrate the bonding of V$_-$ and V$_+$, the debonding of V$_-$-V$_+$, and rebonding of V$_-$ with an adjacent V$_-^*$. The dash-dotted lines represent the boundaries between a positive $B_z$ lobe (marked B$+$) and a negative $B_z$ lobe (marked B$-$).
    (c)(d) are Vortex ring and vortex-droplet ring induced by the same QN of the light source, but with varying $E_0$. The selected light sources are (c) LG$_{02,-1}$ for the star-shaped phase and (d) LG$_{04,1}$ for the flower-shaped phase. The electric field amplitudes $E_1=6.0 A_0 \omega_{GL}$, $E_2=8.0 A_0 \omega_{GL}$, and $E_3=10.0 A_0 \omega_{GL}$ are used to tune the state of the vortex ring. For $E_1$, $E_2$, and $E_3$ the ring consist of 2 vortices, 4 vortices, and vortex droplets, respectively. The rings rotate counterclockwise.}
    \label{fig:multiVP}
\end{figure*}

The formation of vortex ring is shown in~\cref{fig:multiVP}(a). During this process, vortices with the same vorticity ``bond'' together, which initially evolves from ``bonding'' VPs with different vorticities.
The evolution of the vortex ring, as demonstrated in~\cref{fig:multiVP}(b)(c), captures the process of VP generation, where a bonding V$_{-}$ and V$_{+}$ pairs generate, debond, and then rebond again between V$_{-}$ and adjacent V$_{-}^*$. The duration of these bond-debond-rebond process of vortices occurs within $10\tau_{GL}$. Afterward, the ring of vortices stabilizes, with vortices of the same vorticity swirling consistently along the orbital. The tendency for vortices of the same vorticity to group together can be likely attributed to the need to maximize the penetration of $B_z$ through the vortices. After the pairing of V$_{-}$ and V$_{-}^*$ as shown in the right part of~\cref{fig:multiVP}(b), it stably follow the positive $B_z$ lobe. 
This phenomenon occurs in both vortex-ring structures with star-shaped phase profiles (e.g. $s=-1, l \geq 1$ in Supplementary material~\cite{supp2}) and flower-shaped phase profiles ($s=1, l \geq 3$ in Supplementary material~\cite{supp3}). This phenomenon are also found in the data with higher order ($p>0$) of ring-shaped vortices as shown in the Supplementary materials~\cite{supp5,supp6,supp8,supp9}. In the time traces of vortices demonstrated as 3D plot in~\cref{fig:vortex_3D}(d) and~\cref{fig:vortex_3D}(g), this process can be observed at the initial stages of vortex generation as well.

Apart from the bond-debond-rebond process of vortices in the ring structure, we can observe this process in the cases of pure OAM light sources ($l \geq 3, s=0$) as well. In~\cref{app:LP} and Supplementary materials~\cite{supp1}, this process for $s=0$ has starting and ending repetitively, which is unlike the case for the ring-shaped orbital with stable rebonding VPs. The bracket-shaped orbitals provide 4 end points and allow the existence of a transient ``single vortices'' at those end points. One vortex in the rebonding VP with the same vorticity recombines when it touches the vortex at the end point, which is shown in~\cref{fig:LP}.

Moreover, based on the hypothesis that the ring/orbital size influences the formation of vortex droplets and vortices, a ring with more vortices can transform into a ring of vortex droplets under certain conditions.
~\cref{fig:multiVP}(c)(d) demonstrates the progression from a ring of vortices to a ring of vortex droplets as the number of vortices increases, controlled by the amplitude of the electric field.
As the number of vortices grows, the additional vortices follow the orbital path, extending the length of the chain with the same vorticity between the nodes. The nodes maintain the same $n$-fold symmetry in the $\abs{\psi}$ profile. Additionally, the branch cuts arrange in parallel way rather than in series. This arranging preserves the same $n$-fold star and flower symmetries and further results in multi-layered star and flower patterns of $\theta_s$. In~\cref{fig:multiVP}(c)(d), it demonstrates the number of vortices in each chain increases from 2 to 4.

As the electric field increases further, the vortices ``merge'' into a ring of vortex droplets which can be observed in the profiles of both $\abs{\psi}$ and $\omega_{\nu,s}$. In the $\theta_s$ profile, each branch cuts remains distinct and visible. The overall profile remains the ($\abs{l-s}$)-fold symmetric even if the number of vortices increase. Additionally, each droplet in the ring behaves similarly to the braiding vortex droplet, displaying a sharper edge at the front and a trailing tail at the back. In~\cref{app:E}, we demonstrate the corresponding meshgrid to inspect the capability of simulation for spatial resolution of this phenomenon.



\subsection{Radial order $p$}
\label{sec:p}

In LG beams, the QN 
of O, $p$, determines the number of rings in the intensity profile of the electric field. 
Interestingly, it also creates discontinuous phase boundaries for each ring in the phase profile of electric field. When there is no SAM and OAM, the rings exhibit a constant phase, with a $\pi-$phase deviation between adjacent ring, a phenomenon also known as binary-phase~\cite{bencheikh2014generation}. When $l$ is non-zero, the OAM introduces a phase change in the electric field that varies with the azimuthal angle. In this section, we will first examine the effects of $p$ in the context of pure binary-phase and then extend our discussion to include the cases involving $s$ and $l$.


\subsubsection{Imprint effect of binary-phase with $p$}
\label{sec:pure_p}

\begin{figure*}[!htbp]
    \centering
    \includegraphics[width=1\textwidth]{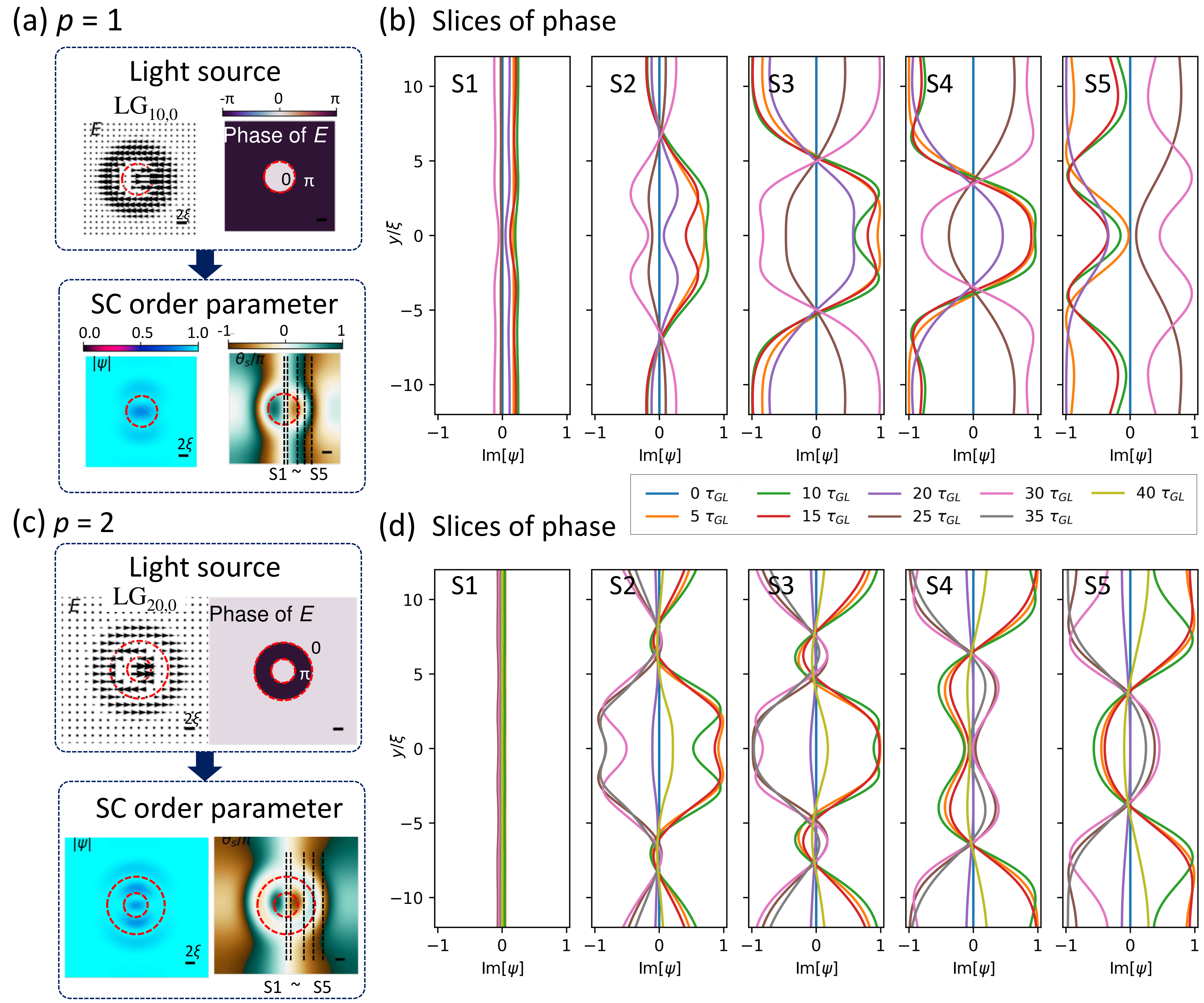}
    \caption{The SC states induced by the $p$-carrying light are presented in this figure. (a) and (b) display the results for $p=1$, while  (c) and (d) correspond to $p=2$. The electric field and phase of $E$ are shown in the ``Light source'' blocks in (a) and (c), while the ``SC order paramter'' blocks illustrate snapshots of $\abs{\psi}$ and $\theta_s$. The red dashed rings indicate the locations of the discontinuity in the phase of $\b{E}$. The black dashed lines labeled S1 to S5 in $\theta_s$ mark the phase cuts used to demonstrate the standing wave in (b) and (d). (b) and (d) show the imaginary part of $\text{Im}[\psi]$ along the cuts from S1 to S5. The different colors of the curves represent various times, as indicated in the legend between (b) and (d).}
    \label{fig:standingwave}
\end{figure*}

In this work, we investigate two cases of binary-phase for the polarization of electric field, specifically for $p=1, 2$. 
In terms of the intensity profile of the electric field, the outer ring generates more linear VPs that are perpendicular to the polarization, as demonstrated in the Supplementary materials~\cite{supp4, supp7}.
Beyond the intensity, the discontinuous phase of the polarization creates circle of singularity (nodal circle), significantly influencing the SC state. The snapshots of the binary-phases are presented in~\cref{fig:standingwave}(a) and (c).

The ring of discontinuity in the phase of the polarization imprints a ring of zero-phase in $\theta_s$ of the SC state via $B_z$ prior to vortex generation. This ring causes the imaginary part of the order parameter, $\text{Im}[\psi]=\abs{\psi} \sin{\theta_s}$, to function like a drumhead for the order parameter.  
In our simulations, clear standing waves are observed at the edge of the ``phase drum''. For the case of $p=2$, two rings of zero-phase in $\theta_s$ manifest. When examining the cuts along the $y$-axis of $\text{Im}[\psi]$, the $\theta_s$ profiles for $p=1$ reveal two nodes of standing wave within the ring, as illustrated in~\cref{fig:standingwave}(b), while $p=2$ exhibits standing waves with four nodes.


\subsubsection{Vortex corral imprinted by $p$ with angular momentums}
\label{sec:p_sl}

When considering $p$ with AMs, the situation becomes more complex. The magnitude of $B_z$ in the inner and outer rings differs significantly. Since formation of vortices strongly depends on the value of the local magnetic flux, vortices in the outer rings might start appearing when the order parameter in the inner ring is fully suppressed. 
We do not investigate a full range of amplitudes of the electric field to study all the possible dynamics in the outer and inner rings, and 
consider here only values  of the amplitude of $\b{E}$ similar to the ones used in the $p=0$ case. The complete set of results for QNs in the range $s=-1 \sim 1$, $l=0 \sim 5$, and $p=0 \sim 2$, along with simulations from 0 to $120\tau_{GL}$, are available in the Supplementary materials~\cite{supp1, supp2, supp3, supp4, supp5, supp6, supp7, supp8, supp9}. Details of the simulations can also be found in ~\cref{app:E}.

\begin{figure*}[!htbp]
    \centering
    \includegraphics[width=1\textwidth]{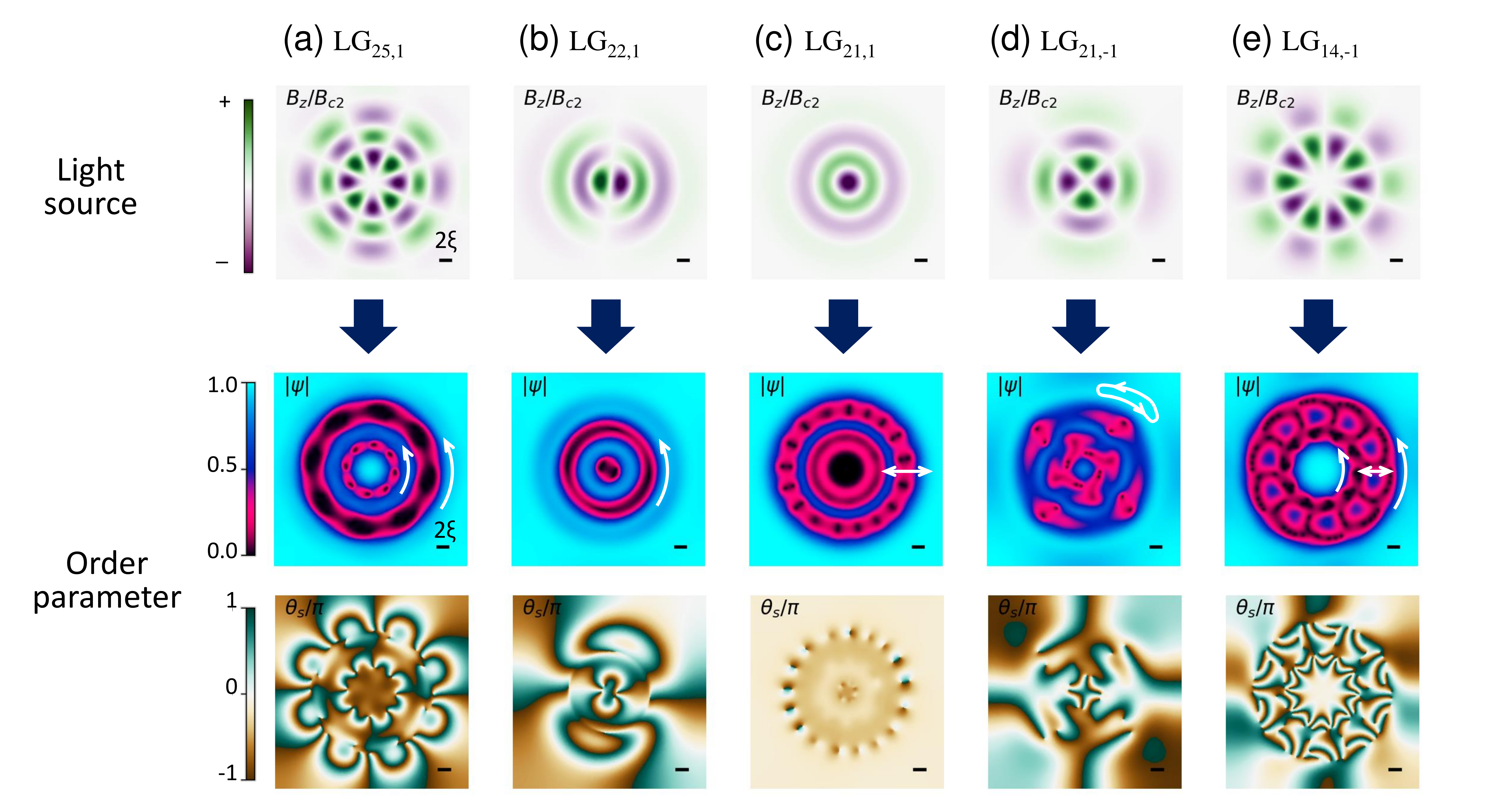}
    \caption{Snapshots of the SC order parameter imprinted by light sources carrying both $p$- and AM. These include: (a) LG$_{25,1}$ showing a multi-ring structure, (b) LG$_{22,1}$ exhibiting large gradient phase induced zero-$\abs{\psi}$ curves, (c) LG$_{21,1}$ inducing a non-rotating vortex-corral, (d) LG$_{25,-1}$ presenting abnormal motion of the outer vortex ring, and (e) LG$_{14,-1}$ showing inter-orbital interactions via vortices.}
    \label{fig:p12}
\end{figure*}

Even with strict constraints on the amplitude of $\b{E}$, the QN of $p$ still introduces new phenomena in the simulations. It leads not only to the appearance of the multiple orbitals of vortices, but to the interactions between orbitals, and results in a lot of kinds of vortex clusters. 
In particular, structured light with non-zero $p$ values generates a larger hollow-ring structure, but still inducing supercurrents and vortices at the center in some cases. This effect resembles an optically induced quantum mirage of superconducting order parameters, similar to the quantum corral effect~\cite{crommie1995quantum}. Consequently, we refer to these vortex clusters as vortex corrals. To illustrate these findings, 
we have selected several noteworthy cases for discussion, which are demonstrated in~\cref{fig:p12}.

\paragraph{Multi-orbital structure} \mbox{}

Firstly, the QN, $p$, is directly linked to the formation of multi-orbital structures. In~\cref{fig:p12}(a), a typical multi-orbital structure is shown, where the inner and outer orbitals share the same symmetry, both exhibiting convex branch cuts. This example also clearly illustrates the chain of vortex droplets. Across various QN configurations, this phenomenon occurs more readily when ($s,l,p$)=(1,$\geq$2,2), resulting in a multi-flower-shaped phase, as demonstrated in~\cref{fig:slp} and in the Supplementary material~\cite{supp9}.

In the absence of SAM, for QNs $l>1$, $s=0$, and $p=0$, for which the results are shown in~\cref{sec:l>1_OAM}, the pure OAM effect leads to stable, bracket-like structures when $l \geq 3$. However, when combined with the OR, $p$, the results become more complex, sometimes producing exotic patterns such as a digital ``8'' profile for ($s,l,p$)=(0,2,2) and ``X'' profile for ($s,l,p$)=(0,1,2). In the most of the cases, especially for $l \geq 3$, the structures form multi-layered bracket patterns oriented perpendicularly to each other. These outcomes are presented in~\cref{fig:slp} and Supplementary materials~\cite{supp1,supp4,supp5}.

\paragraph{Zero-$\abs{\psi}$ curves} \mbox{}

In the case shown in~\cref{fig:p12}(b), curvy voids (regions of suppressed amplitude of the order parameter) appears in the $\abs{\psi}$ profile. In the phase profile, there is a distinguishable contour of circular shape that corresponds to these voids. It is formed by a fast change in the phase gradient. The data suggest that across certain segments of this contour the amplitude of the gradient is very large or there might be a discontinuity in phase value. 
If such a discontinuity exists, its origin remains unclear and needs further investigation. The contour also contains four singularity points around which the phase winds by $2\pi$, which points to the presence of vortices in the voids.
This contour remains stable and rotates with the light. 
This phenomenon is also frequently observed when ($s,l,p$)=(1,$\geq$1,2) , exhibiting a multi-flower-shaped phase, as demonstrated in ~\cref{fig:slp} and Supplementary materials~\cite{supp6,supp9}.

\paragraph{Non-rotating vortex corral} \mbox{}

With higher RO, $p$, and $\abs{l-s}=0$ the motion of vortices occur in radial direction resembles the case $\abs{l-s}=0$ for $p=0$ described in~\cref{sec:sl_OAM}. In this case, the profile of magnetic field do not rotate which results in linear motion of the vortices in the VPs. 
The resulting configuration of VPs resembles a corral structure, which we call a VP-corral.
The example in~\cref{fig:p12}(c) also shows that VP generation occurs in the vicinity of the first and third nodal lines of $B_z$, but not in the vicinity of the second nodal line of $B_z$.
Similar dynamics of vortices occurs for the light beams LG$_{11,1}$ and LG$_{21,1}$, as demonstrated in the Supplementary materials~\cite{supp6,supp9}.

\paragraph{Abnormal motion of the outer vortex ring}\mbox{}

For $p=2$, the outer ring is close to boundary. When $s=-1$, it shows the abnormal motion of the outer vortex ring unlike for the case of $s=1$, and other vortex rings when $p=0, 1$. The VPs are generated at diagonals of the sample, which for $l=2$ does not respect the symmetry of $B_z$. Besides, those VPs are trapped at the same locations recombining shortly after the generation, at least for the period of time investigated in the simulation. It can be attributed to the closeness of the outer ring to the boundary. This phenomenon can also be observed in other cases, such as when ($s,l,p$)=(-1,$\geq$2,2), though it is less pronounced compared to the LG$_{21,-1}$ case.

\paragraph{Inter-orbital interaction of vortices} \mbox{}

The previous cases exhibit various multi-orbital behaviors, but there is no direct connection between the orbitals. However, with higher values of $p$ and $l$, and sufficient applied electric field amplitude, orbitals can become interconnected, forming a vortex cluster similar to a snowflake. ~\cref{fig:p12}(e) shows a typical example of a ``vortex flake'' induced by LG$_{14,-1}$. In this case, aside from the vortices aligned with the rotational direction, there are also VPs oriented transversely along the ring. This structure allows for the coexistence of both rotating and linear vortex motions, which could offer new insights into the hydrodynamics of superfluids.


\subsection{Diagram of $s$, $l$, and $p$}
\label{sec:QNs}

\begin{figure}[!htbp]
    \centering
    \includegraphics[width=0.5\textwidth]
    {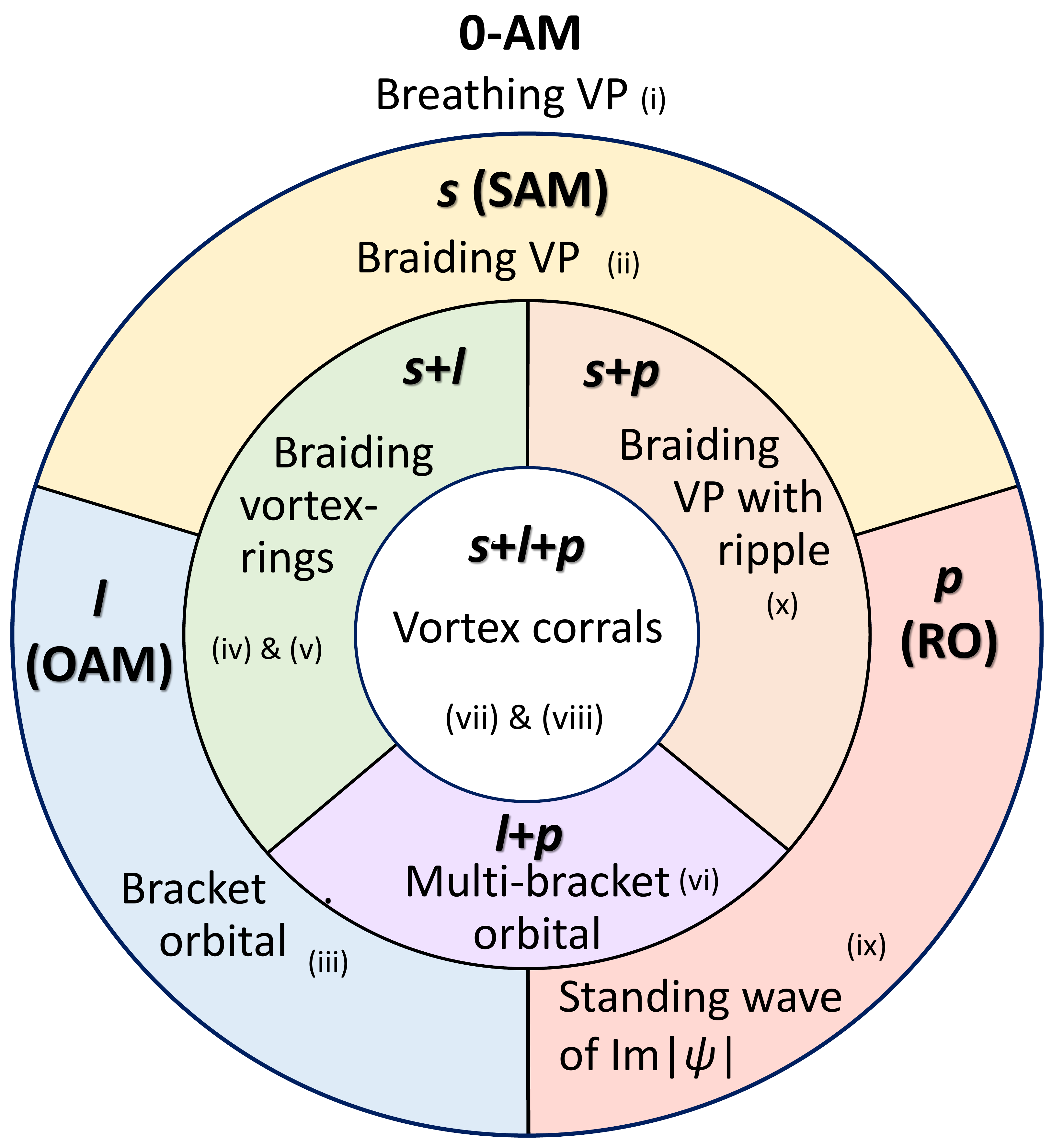}
    \caption{Schematics QN diagram of SC state induced by LG beam.
    The diagram illustrates the sets and subsets of quantum numbers $s$, $l$, and $p$ for SAM, OAM, and RO. The details are shown in~\cref{tab:phase_diag}.
    }
    \label{fig:phase_diag}
\end{figure}

Considering the QNs $s$, $l$, and $p$, the QN diagram can be represented as a collection of QN sets and their union of subsets, as illustrated in ~\cref{fig:phase_diag}. From this QN diagram, the order parameter dynamics induced by structured light can be categorized into the subsets mentioned in~\cref{tab:phase_diag} (with $l > 0$ as an example, where the sign of SAM determines whether the rotation aligns with or opposes the OAM). Besides, the snapshots and dynamical simulation results are provided in~\cref{app:psi} and Supplementary materials~\cite{supp1, supp2, supp3, supp4, supp5, supp6, supp7, supp8, supp9}.

\begin{table*}[!htbp]
    \caption{Details of the QN diagram shown in ~\cref{fig:phase_diag}. Blue and yellow dots represent the vortices with different vorticity. Black dots represent clusters of vortices or vortex droplets. Red dashed curve indicate the rotational direction. Green and purple curves in (ix) depict the curve of $\theta_s=0$ and $Im[\psi]$, respectively. Since some of the vortex motions are too complicate or no clear vortex, so we use pink lines to represent orbitals for those situations. } \label{tab:phase_diag}
  \begin{tabularx}{1\linewidth}
      {lp{1in}X} 
    \hline 
    No. & Schematic& Description  \\
      \hline 
      (i) & \parbox[c]{1em}{\includegraphics[width=1.1in]{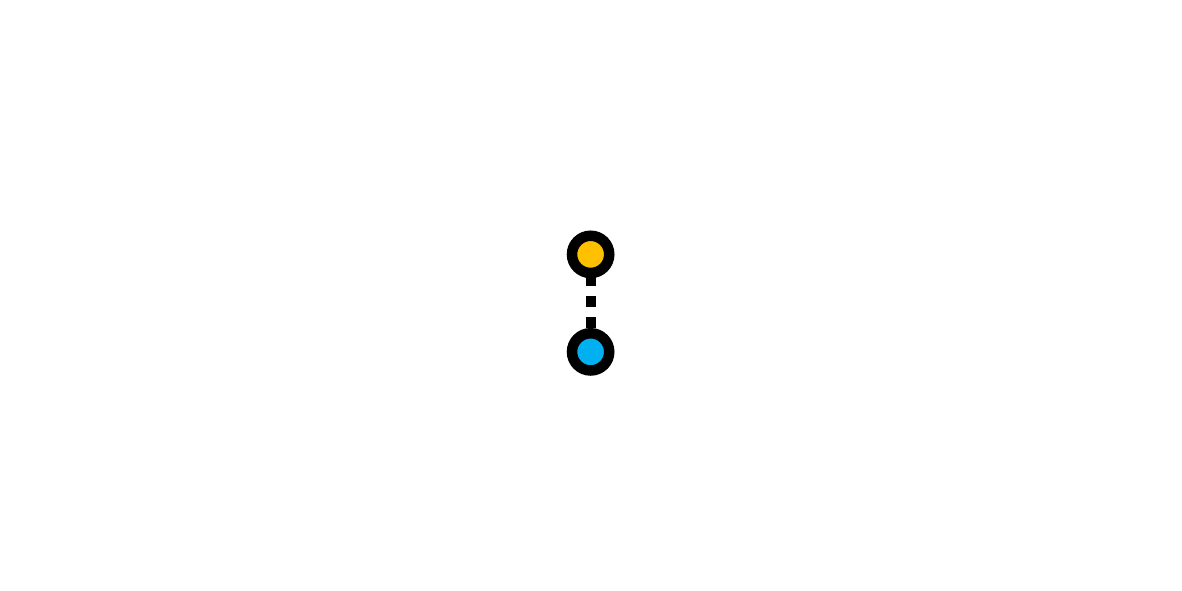}} 
      & No AM: $(s,l,p)=(0,0,0)$.\newline
        The LP Gaussian light beam has no angular momentum. The VP(s) exhibit linear or breathing-like motion, moving perpendicular to the polarization direction.  \\  
      (ii) & \parbox[c]{1em}{\includegraphics[width=1.1in]{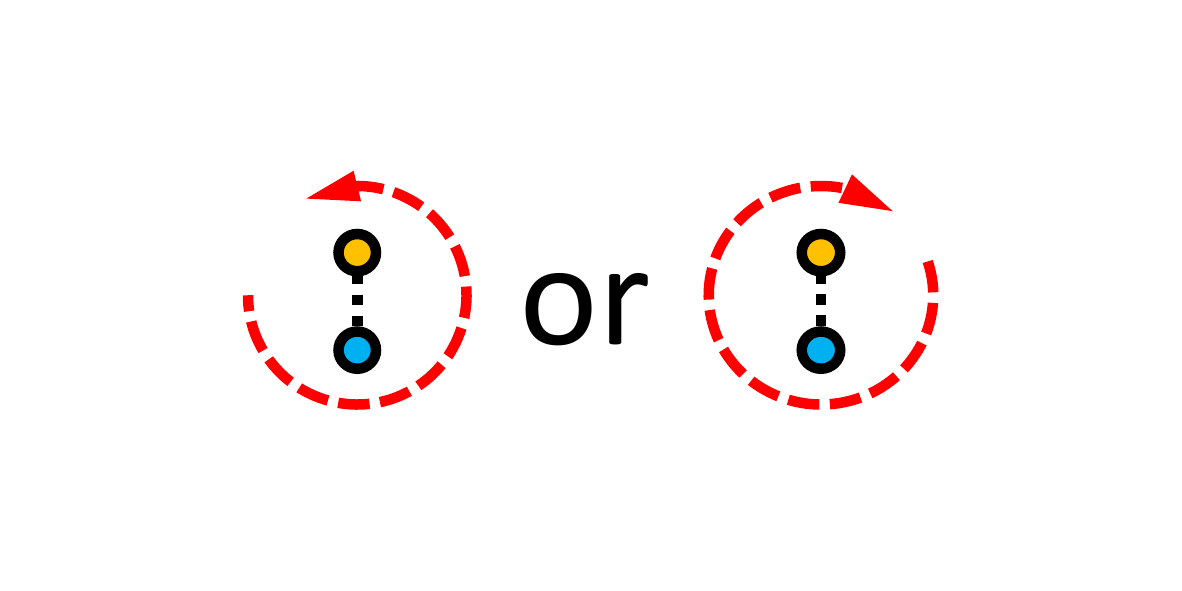}} 
      & $\pm$SAM: $(s,l,p)=(\pm1,0,0)$.\newline
        Right- and left-handed CP Gaussian light beams are determined by $s=-1$ and $+1$, respectively. Braiding VP(s) or braiding droplets rotate clockwise or counterclockwise around the center of the spot.  \\  
      (iii) & \parbox[c]{1em}{\includegraphics[width=1.1in]{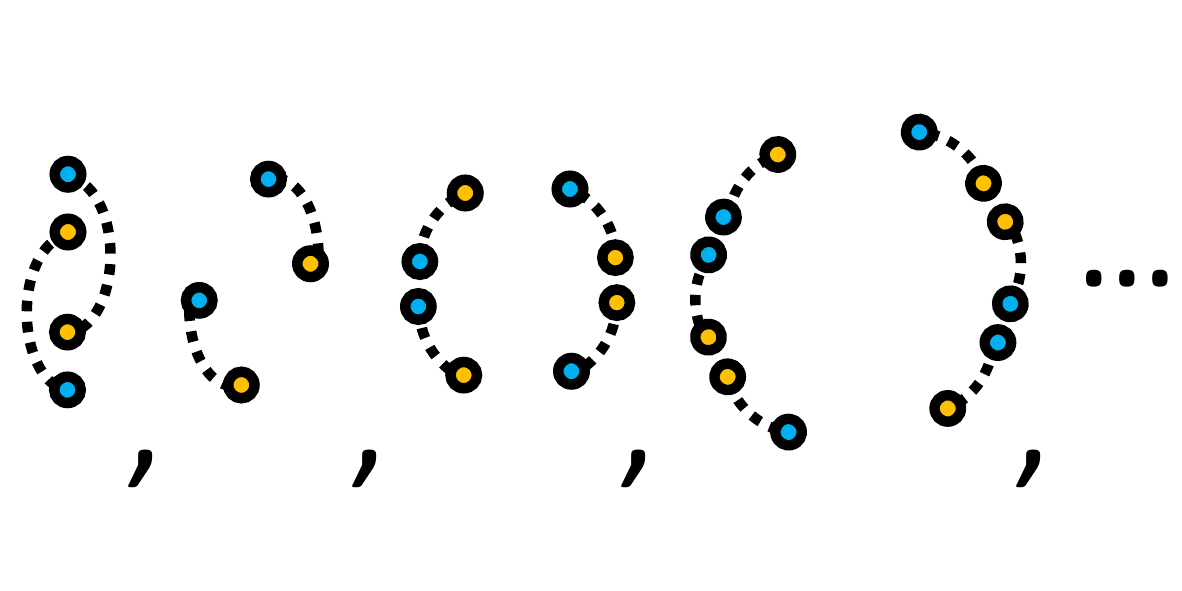}} 
      & OAM: $(s,l,p)=(0,>0,0)$.\newline
        The LP LG light beam revolves in the right-handed direction. The dynamics of  $\abs{\psi}$ shows $C_2$ symmetry, and vortices move along the orbitals that have shape of round brackets when $l \geq 3$.  \\  
      (iv) & \parbox[c]{1em}{\includegraphics[width=1.1in]{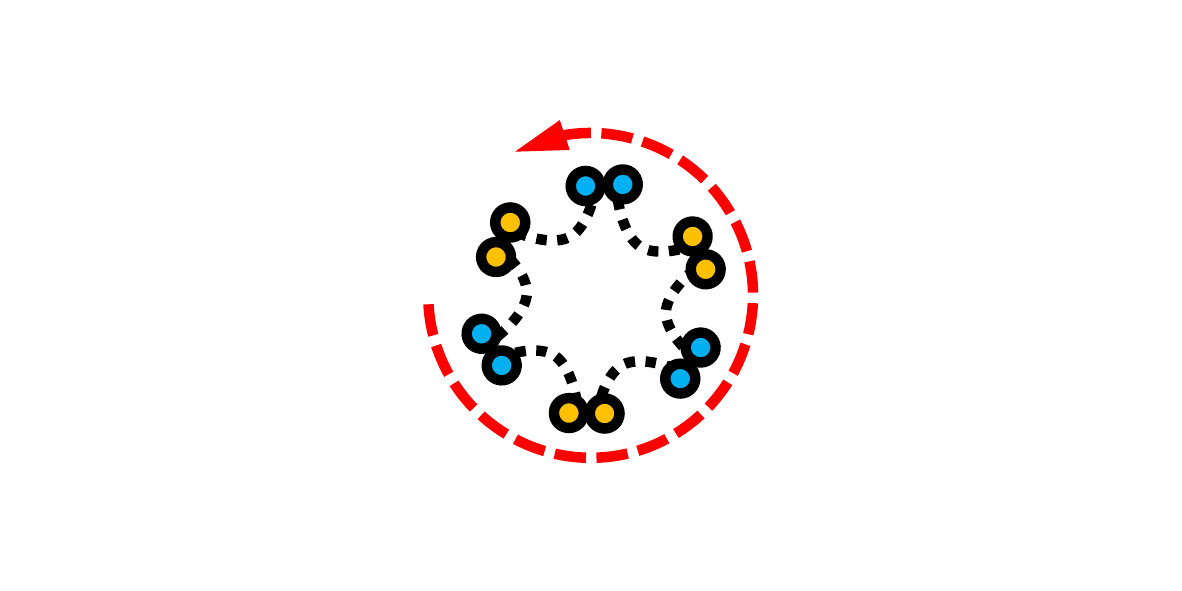}} 
      & OAM$+$SAM: $(s,l,p)=(-1,>0,0)$.\newline
        These LG light beams generate a $2\abs{l-s}$-fold symmetric vortex ring in the $\abs{\psi}$ profile, and $\abs{l-s}$-fold symmetries in both $\theta_s$ and $\omega_{\nu,s}$. The beams produce a star-shaped $\theta_s$, which rotates in the right-handed direction.  \\  
      (v) & \parbox[c]{1em}{\includegraphics[width=1.1in]{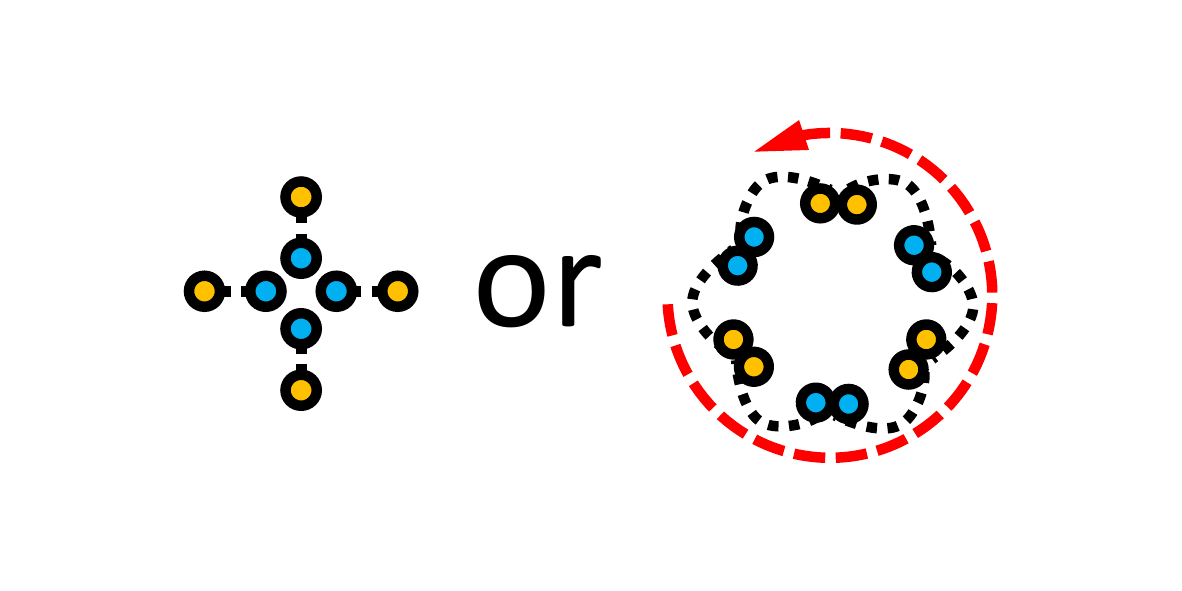}} 
      & OAM$-$SAM: $(s,l,p)=(1,>0,0)$. \newline
        Similar to (iv), but produce a flower-shaped phase $\theta_s$. For $s = l = 1$, the magnetic field becomes rotationally symmetric, which results in linear motion of the vortices.  \\  
      (vi) & \parbox[c]{1em}{\includegraphics[width=1.1in]{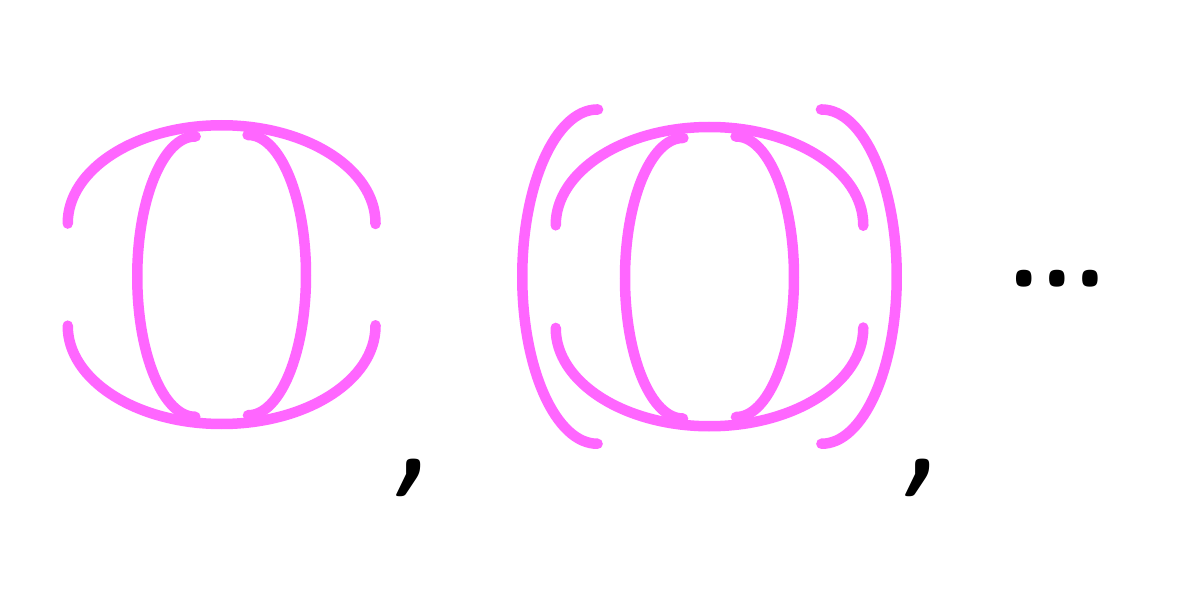}} 
      & OAM \& RO: $(s,l,p)=(0,>0,>0)$.\newline
        The LP LG light beam develops nodal lines in the electric field and $z$-component of the magnetic field. 
        For $l\geq3$, these $B_z$ results in multiple regions of suppressed $\abs{\psi}$, which have shapes of round brackets, in which the vortices move. For $l<3$, these $B_z$ result in other types of orbitals along which the vortices move. \\
      (vii) & \parbox[c]{1em}{\includegraphics[width=1.1in]{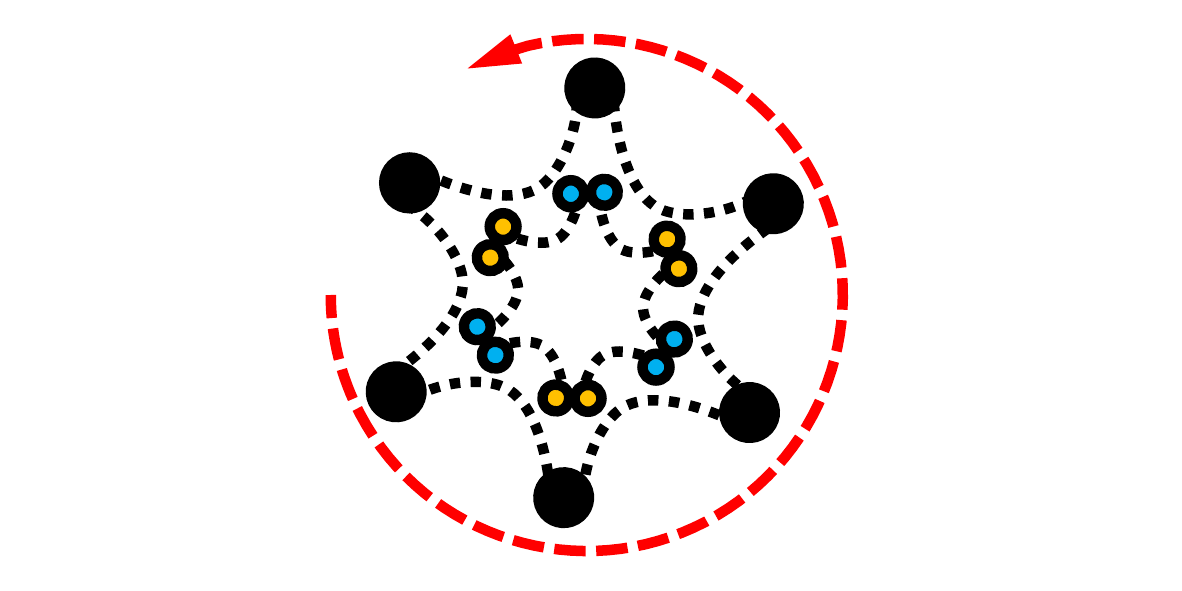}} 
      & OAM$+$SAM \& RO: $(s,l,p)=(-1,>0,>0)$. \newline
        This LG light beam forms co-axial multi-vortex rings with $\abs{l-s}$-fold symmetries in both $\theta_s$ and $\omega_{\nu,s}$, and $2\abs{l-s}$-fold symmetry in $\abs{\psi}$. It generates complex vortex clusters, or so-called vortex corrals, accompanied by star-shaped set of branch cuts in $\theta_s$ that rotate in the right-handed direction. According to our tests, for $p = 1$ and $2$, the results of simulations display single and double vortex rings, sometimes influenced by boundary interactions or accompanied by inter-ring dynamics. \\  
      (viii) & \parbox[c]{1em}{\includegraphics[width=1.1in]{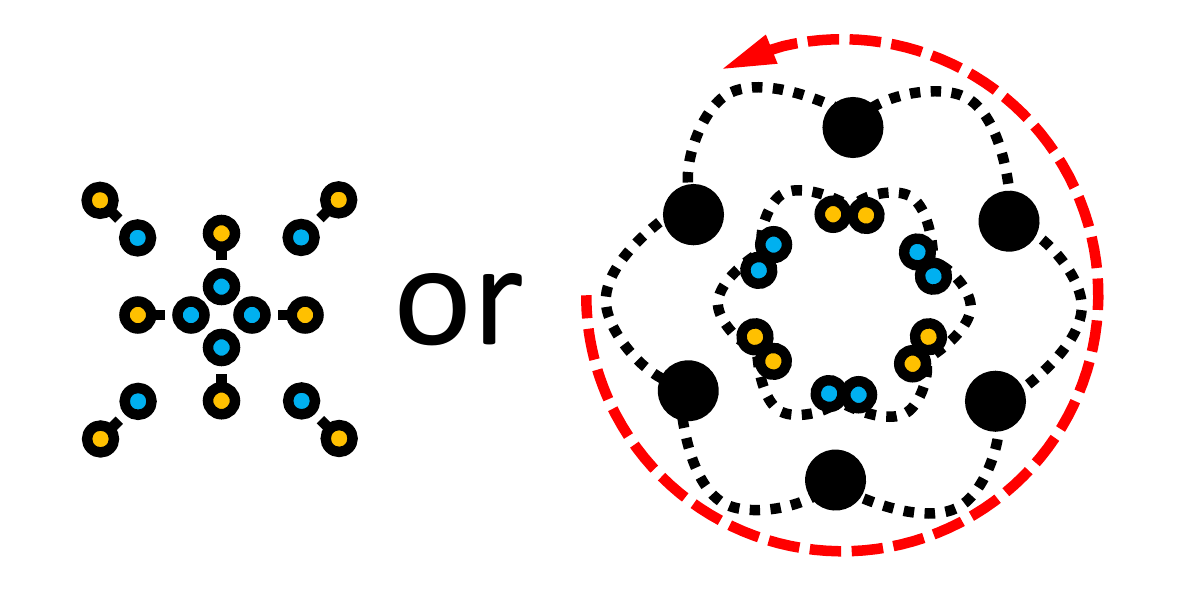}} 
      & OAM$-$SAM \& RO: $(s,l,p)=(1,>0,>0)$.\newline
        Similar to (vii), but with flower-shaped set of branch cuts in $\theta_s$. For $s = l = 1$, it results in linear motion of the vortices. For $p=1,2$, the dynamics become increasingly intricate, with inter-ring interactions and large gradients of the phase in the vicinity of zero-$\abs{\psi}$ curves.  \\  
      (ix) & \parbox[c]{1em}{\includegraphics[width=1.1in]{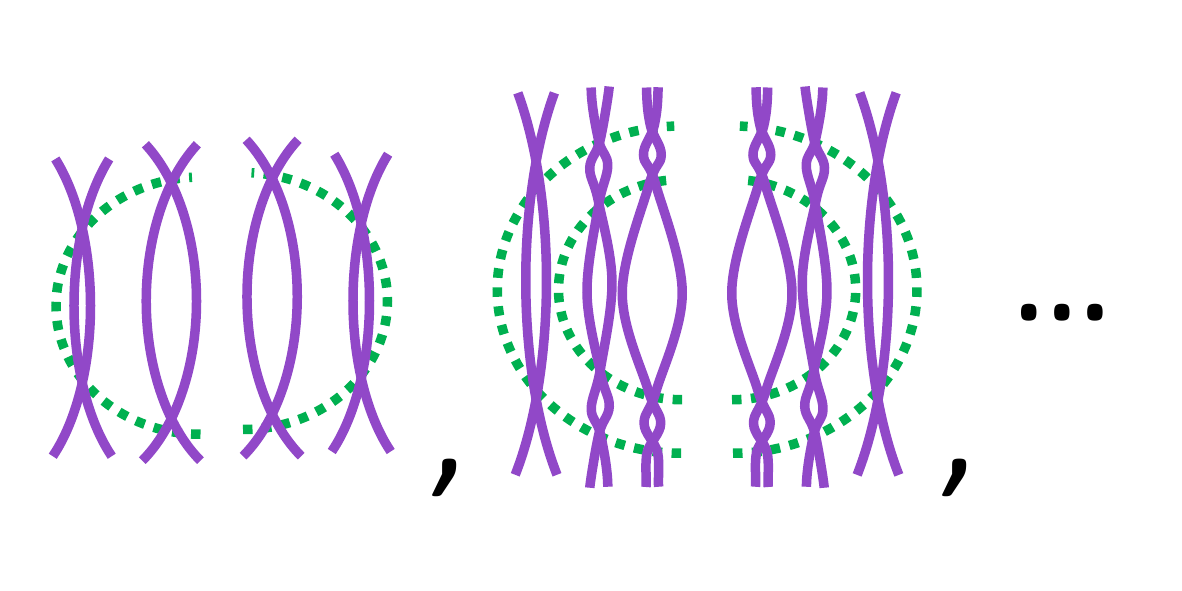}} 
      & RO: $(s,l,p)=(0,0,>0)$. \newline
        The LG light beam exhibits binary-phase polarization. The order parameter initially presents standing waves in the imaginary component of $\psi$, accompanied by ripples in $\abs{\psi}$ before vortex formation. Once the vortices emerge, the dynamics of $\abs{\psi}$ exhibits linear motion of the vortices in the VPs.  \\  
      (x) & \parbox[c]{1em}{\includegraphics[width=1.1in]{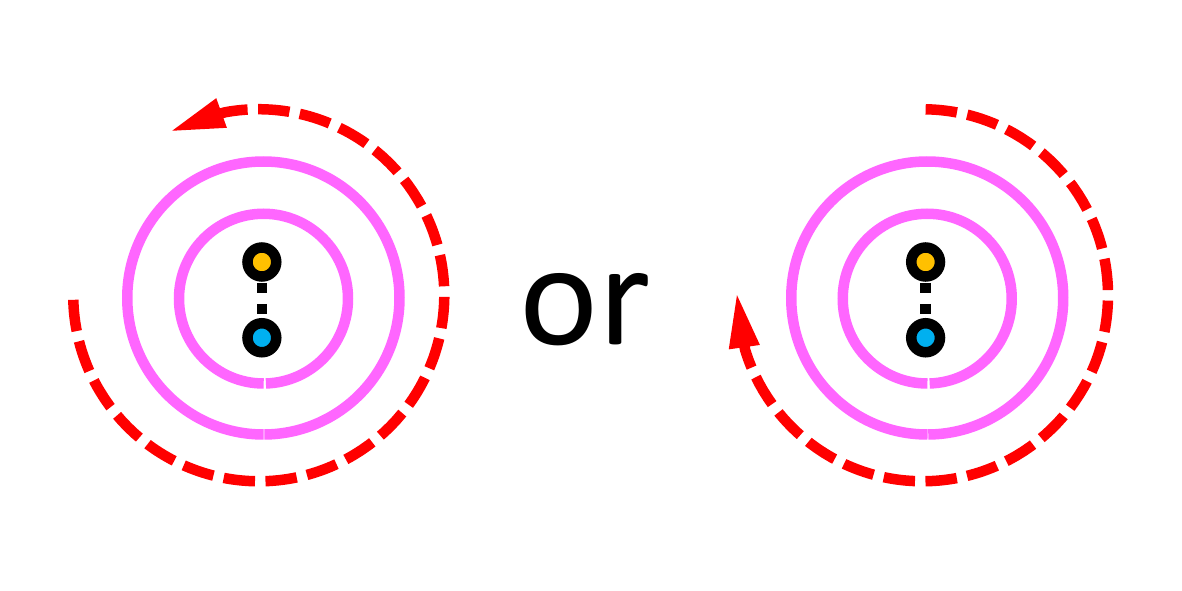}} 
      & $\pm$SAM \& RO: $(s,l,p)=(\pm1,0,>0)$.\newline
        Braiding VP(s) or braiding droplets of vortices rotate right-handedly ($s=-1$) or left-handedly ($s=+1$) around the center of the beam, accompanied by ripple patterns in $\abs{\psi}$.  \\ 
      \hline
  \end{tabularx}
\end{table*}


\section{\label{sec:summary}Conclusion}

In this work, we demonstrate the {\em quantum printing } of structured light on superconductors, where the vorticity of the vector potential of light is transferred to the vorticity of the SC supercurrent and vortex motion. By tuning the QNs of LG beams — including SAM, OAM, and RO — and adjusting the amplitude of the electric field, we can steer and manipulate 
superconducting vortices.

We demonstrated a variety of vortex dynamics
including breathing VPs (e.g.,~\cref{fig:vortex_3D}(a)), braiding VPs (e.g.,~\cref{fig:vortex_3D}(b)), vortex droplets (e.g.,~\cref{fig:multiVP}(c)(d)), optical angular momentum induced linear VPs motion (e.g.,~\cref{fig:phase_front_shear}(d)), and swirling vortex rings with star- and flower-shaped phase profiles (e.g.,~\cref{fig:vortex_3D}(d)(g)). We also observe more complex phenomena of vortex corrals such as vortex-flake structures (e.g.,~\cref{fig:p12}(e)), multi-ringed vortex configurations (e.g.,~\cref{fig:p12}(a)(b)), standing wave patterns (e.g.,~\cref{fig:standingwave}), and asymmetric orbital behavior 
(e.g.,~\cref{fig:sl}(a)). These phenomena reflect the direct imprint of light’s QNs onto the SC system, effectively imprinting the structured light's symmetry and dynamics to the superconducting condensate.

Through simulations, we show how manipulating the QNs of the light beam allows for sophisticated control over vortex formation and motion, which offers potential applications in fields like quantum hydrodynamics and optical manipulation of SC states. 
Experimentally observable sugnatures of the phenomena we present  have been discussed in previous work~\cite{LG_TDGL_I}. We note the experimental observation of this rich set of vortex dynamics will require coupling structured THz light to a superconducting thin film while taking steps to avoid undesired heating and quasiparticle excitation of the film. More specific examples of detection would depend on the expeimental platform used to detect vortex motion. We mention the I(V) characteristics, local scanning tunneling microscope (STM) probes and perhaps the angle-resolved photoemission spectroscopy (ARPES) in the time domain as possible routes to probe the predicted effects. 

The diagram of QNs in~\cref{fig:phase_diag} 
summarizes the collective response of SC order parameter to different combinations of QNs carried by light. It demonstrates  how structured light serves as an optical ``stamp,'' being able to imprint its symmetry onto the superconductor and influence vortex dynamics. The results pave the way for further exploration of light-superconducting interactions, with potential implications for controlling topological defects, 
vortex sheet~\cite{volovik2015superfluids,thuneberg1995introduction}, optical quantum switches, SC devices~\cite{yerzhakov2024induction,dremov2019local}, and other phenomena in condensed matter systems.


\section{\label{sec:acknowledg}Acknowledgments}

Part of this work was carried out while AB and his collaborators were visiting Stanford Department of Physics and SLAC. We are grateful to T. Deveraux and members of his group for discussions and hospitality during our visit.  We also acknowledge useful discussoins with S. Bonetti K. Bondarenko, A. Chainani, B. Friedman,  G. Jotzu, J. T. Heath , H. Hwang, R.B. Laughlin, D.H. Lee,  W. Lee, M. Manson, A. Metha, A. Pathapati,  Z.X Shen, Y. Suzuki, C.L.C. Triola,  O. Tjernberg, V. Unikandanunni J. Wiesenrieder,  P. Wong,and  M. Verma.

This work was supported by the US  DOE BES DE-SC0025580 (AB and TY) and the European Research Council under the European Union Seventh Framework ERC-2018-SYG 810451 HERO (HY).

\appendix
\begin{appendices}

\section{\label{app:Bz}Contribution of $r$ and $\varphi$ to $B_z$} 

\begin{figure*}[!htbp]
    \centering
    \includegraphics[width=.8\textwidth]{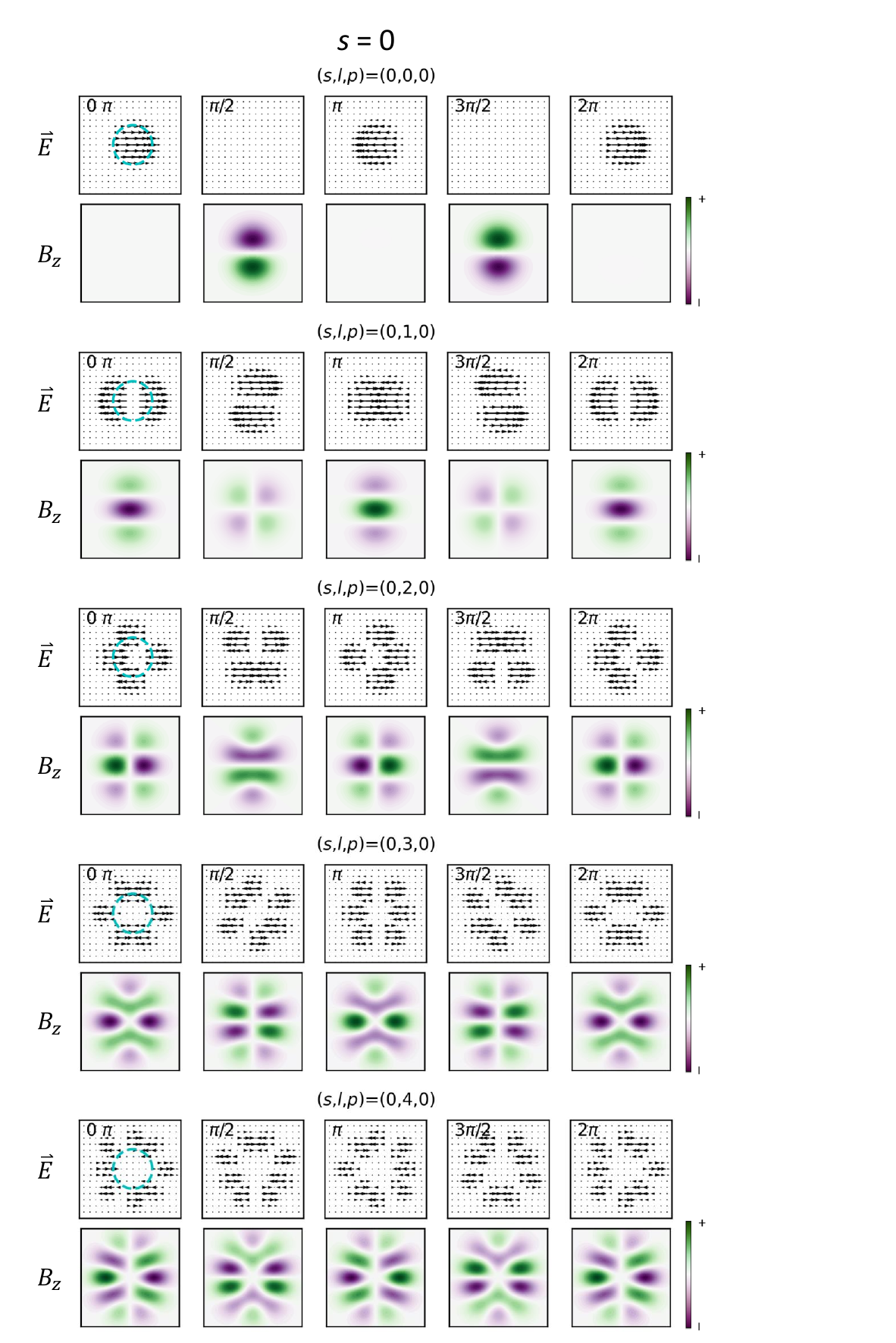}
    \caption{Snapshots of normalized $E$ and normalized $B_z$ at $\omega_{EM}t=0,\pi/2, \pi, 3\pi/2,2\pi$} for each  of the light sources labeled ($s=0,l,p$). The dashed blue circles mark the spot size.
    \label{fig:Bz_snapshot1}
\end{figure*}

\begin{figure*}[!htbp]
    \centering
    \includegraphics[width=.8\textwidth]{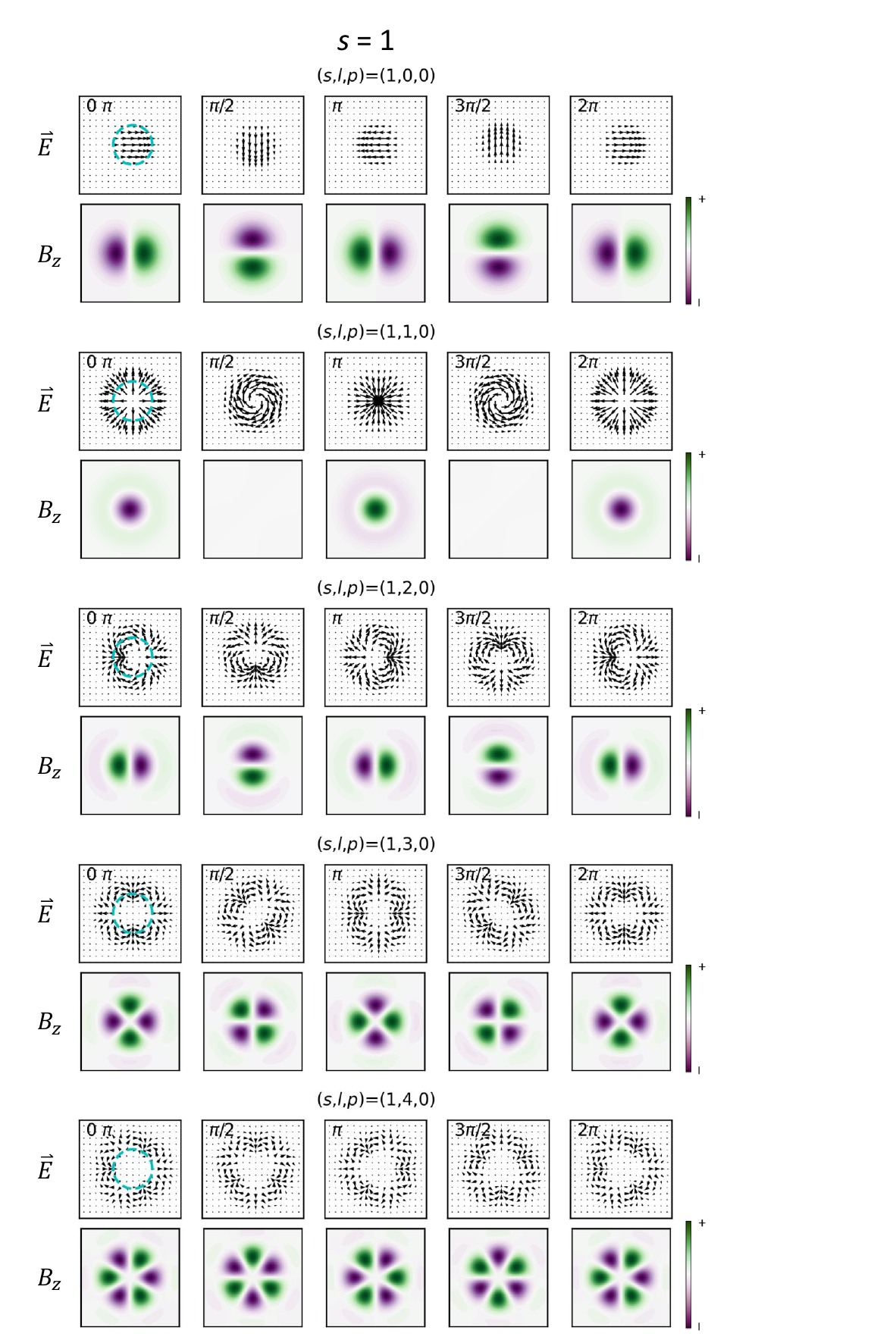}
    \caption{Snapshots of normalized $E$ and normalized $B_z$ at $\omega_{EM}t=0,\pi/2, \pi, 3\pi/2,2\pi$} for each of thelight sources labeled ($s=1,l,p$). The dashed blue circles mark the spot size.
    \label{fig:Bz_snapshot2}
\end{figure*}

\begin{figure*}[!htbp]
    \centering
    \includegraphics[width=.8\textwidth]{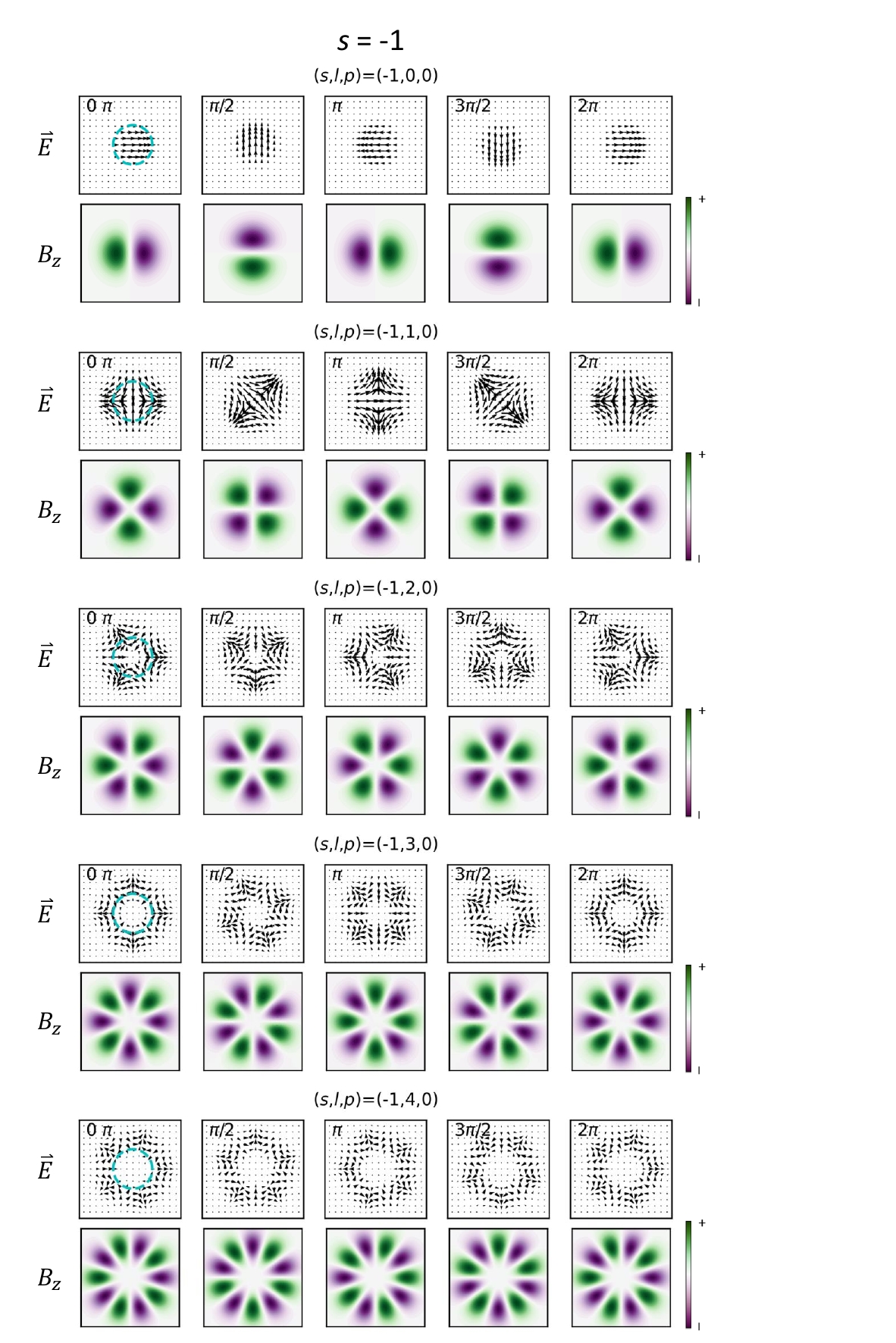}
    \caption{Snapshots of normalized $E$ and normalized $B_z$ at $\omega_{EM}t=0,\pi/2, \pi, 3\pi/2,2\pi$} for each  of the light sources labeled ($s=-1,l,p$). The dashed blue circles mark the spot size.
    \label{fig:Bz_snapshot3}
\end{figure*}

For the LG beam, $z$-component of the magnetic field, $B_z=(\curl{\b{A}})_z$, can be non zero.
The amplitude of the electric field in~\cref{eq:LG_u} consists of three components,
\begin{equation} 
\label{eq:LG_u0*ur*uphi}
u_{p,l}\left(r,\varphi,z\right)= u_0 u_r u_{\varphi},
\end{equation}
where $u_0$, $u_r$, $u_{\varphi}$ are the constant of amplitude equal to $E_0$, radial distribution of amplitude, and azimuthal distribution, respectively.
When the light sources with constant $u_{\varphi}$, such as LG$_{p0,s}$ for any $s,p$, the curl of transverse mode becomes the curl of amplitude distribution as $\curl{(u_0 u_r u_{\varphi}\hat{\b{E}}_{pol})} = u_0 u_{\varphi} \curl{(u_r\hat{\b{E}}_{pol})}$, where $\hat{\b{E}}_{pol}=\cos{\varphi_{xy}}\hat{\mathbf{e}}_{x}+e^{-i \sigma} \sin{\varphi_{xy}}\hat{\mathbf{e}}_{y}$ represents the unit vector for polarization based on~\cref{eq:LG_E}.
Hence, it is completely dominated by the distribution of intensity.
If we consider the OAM with the azimuthal term $\exp{i l \varphi}$ where azimutal angle can be expressed as $\varphi=\arctan (y/x)$, then the derivative of $\varphi$ becomes non-negligible. The results of numerical calculations are provided in Figs. \ref{fig:Bz_snapshot1}-~\ref{fig:Bz_snapshot3}.

\section{\label{app:E} Simulation inputs: $E$, $L$ and mesh} 

In this Appendix, we provide information on the amplitude of the electric field $E_0$ in~\cref{eq:LG_u} (see~\cref{tab:E0}), the sample size $L$ (see~\cref{tab:L}), and the mesh we used in the simulations.
The electric field conditions vary across simulations for three main reasons:
(1) The peak intensity of LG beams is not solely determined by $E_0$. For example, OAM-carrying beams, which have hollow centers, exhibit lower peak amplitudes compared to Gaussian beams with the same $E_0$.
(2) The resulting $B_z$ field, derived from the curl of the vector potential, depends on the specific structure of the light.
(3) Geometry of the sample and the symmetry of the resulting $B_z$ may influence the minimum threshold for a VP generation.
Table~\ref{tab:E0} lists the $E_0$ values used in all simulations unless otherwise specified in the figure captions.

\begin{table}[!htbp]
\caption{The list of $E_0$ in the units of $A_{0} \cdot \omega_{GL}$.}
\begin{tabular}{cccccccccc}
\hline
$p$      & \multicolumn{3}{c}{0} & \multicolumn{3}{c}{1} & \multicolumn{3}{c}{2} \\ \hline
\backslashbox{$l$}{$s$} & -1     & 0     & 1    & -1     & 0     & 1    & -1     & 0     & 1    \\ \hline
0        & 2.0    & 2.0   & 2.0  & 2.0    & 2.0   & 2.0  & 3.0    & 2.0   & 3.0  \\
1        & 2.5    & 3.0   & 4.0  & 4.0    & 3.0   & 4.0  & 5.0    & 4.0   & 6.0  \\
2        & 6.0    & 5.0   & 3.0  & 3.5    & 4.0   & 3.5  & 5.0    & 5.0   & 4.0  \\
3        & 8.0    & 5.0   & 5.0  & 3.5    & 3.5   & 3.5  & 5.0    & 4.0   & 4.0  \\
4        & 8.0    & 6.0   & 6.0  & 8.0    & 5.0   & 4.0  & 8.0    & 5.0   & 4.0  \\
5        & 8.0    & 7.0   & 7.0  & 8.0    & 6.0   & 6.0  & 8.0    & 8.0   & 8.0  \\ 
\hline
\end{tabular}
\label{tab:E0}
\end{table}

\begin{table}[!htbp]
\caption{The list of $L$ in the units of $\xi$.}
\begin{tabular}{cccccccccc}
\hline
$p$      & \multicolumn{3}{c}{0} & \multicolumn{3}{c}{1} & \multicolumn{3}{c}{2} \\ \hline
\backslashbox{$l$}{$s$} & -1     & 0     & 1    & -1     & 0     & 1    & -1     & 0     & 1    \\ \hline
0        & 25     & 25    & 25   & 30     & 30    & 30   & 40     & 40    & 40   \\
1        & 25     & 25    & 25   & 30     & 30    & 30   & 40     & 40    & 40   \\
2        & 25     & 25    & 25   & 30     & 30    & 30   & 40     & 40    & 40   \\
3        & 25     & 25    & 25   & 30     & 30    & 30   & 40     & 40    & 40   \\
4        & 30     & 25    & 25   & 40     & 30    & 30   & 40     & 40    & 40   \\
5        & 30     & 25    & 25   & 40     & 30    & 30   & 40     & 40    & 40   \\ \hline
\end{tabular}
\label{tab:L}
\end{table}

\begin{figure*}[!htbp]
    \centering
    \includegraphics[width=0.9\textwidth]{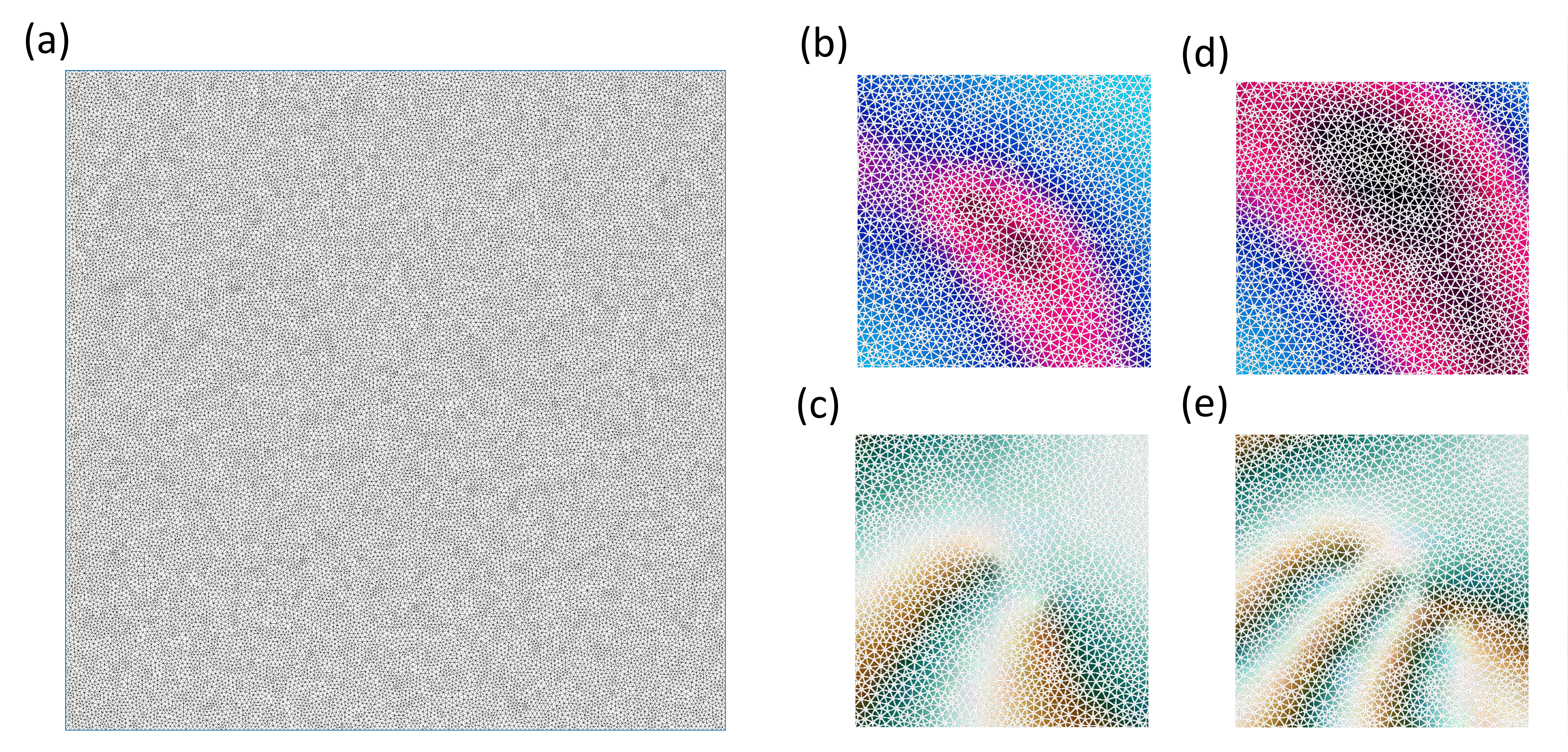}
    \caption{(a) The mesh of the case $L=25 \xi$, as black lines. (b)-(d): The meshgrids (white lines) overlap with $\abs{\psi}$ in (b)(d) and $\theta_s$ in (c)(e). The order parameters are the snapshots from~\cref{fig:multiVP}(c). Figs (b)(c) shows the case of vortices, which correspond to $E_0=E_1$ in~\cref{fig:multiVP}(c), and Figs. (d)(e) The case of the droplet from $E_0=E_3$ in~\cref{fig:multiVP}(c).}
    \label{fig:mesh}
\end{figure*}

The step size of the mesh should be smaller than the unit of length, $\xi$, in order to resolve the vortices. Even if the morphology of vortex is not the topic of this work, however, the scale of the size remains important because sometimes the vortices are too crowded and therefore merge to vortex droplets. The step size need to be sufficiently small to support this result. Here, we demonstrate the mesh for the case $L=25\xi$ in~\cref{fig:mesh}(a). We also overlap the meshgrid with the vortices and the vortex droplet from the snapshots in~\cref{fig:multiVP}(c). Apparently, in the case of the droplet (~\cref{fig:mesh}(e)), which represents vortices close to each other, the meshgrid still can spatially resolve the singularity points of $\theta_s$. 

\section{\label{app:vorticity} Imprint effect of vorticity} 

\begin{figure*}
    \centering
    \includegraphics[width=1\textwidth]{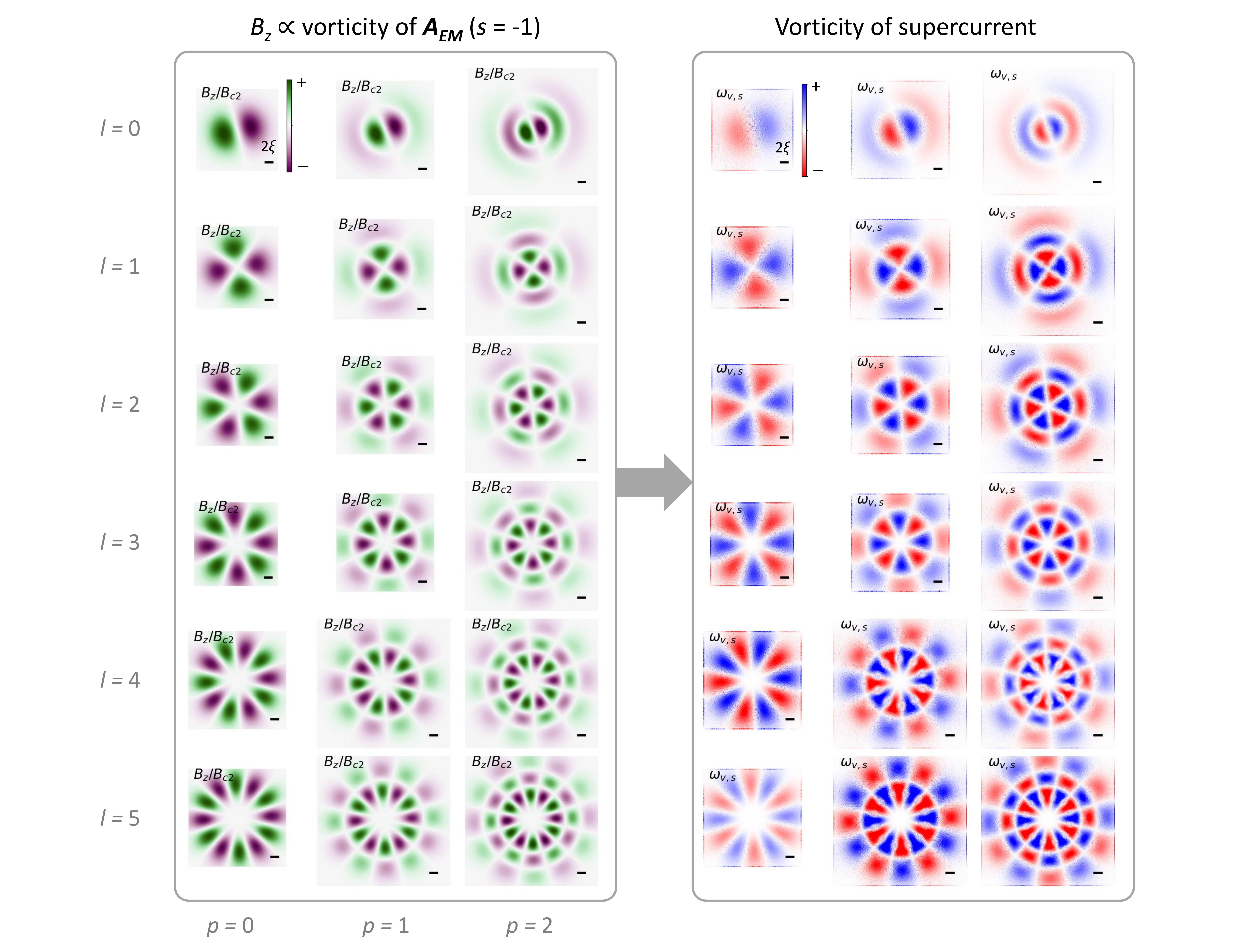}
    \caption{The vorticity of vector potential and imprinted vorticity of supercurrent before vortex generation. The light sources are LG$_{pl,-1}$. All of profiles are snapshots at $20 \tau_{GL}$.} 
    \label{fig:Bz_Wn1_A1}
\end{figure*}

\begin{figure*}
    \centering
    \includegraphics[width=1\textwidth]{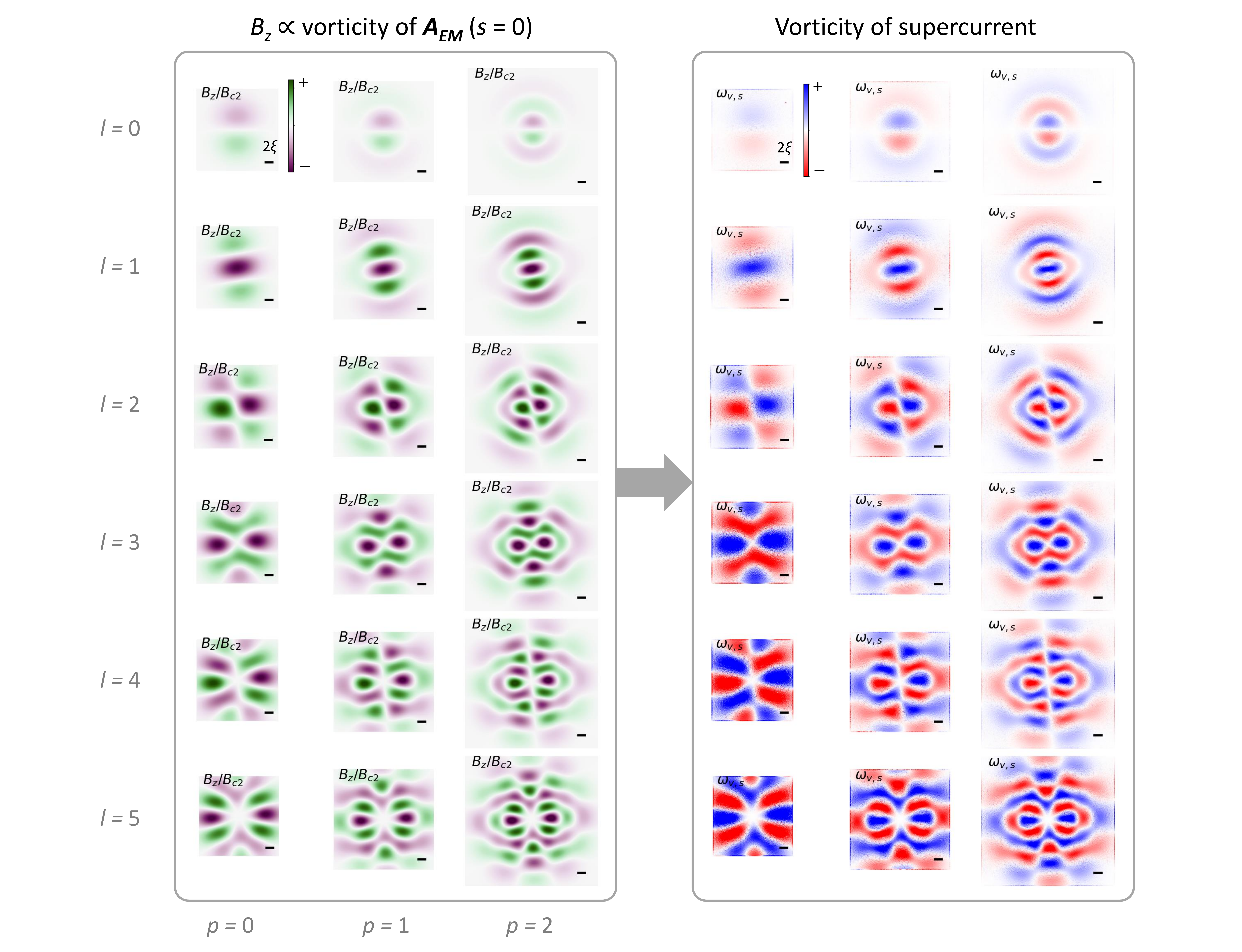}
    \caption{The vorticity of vector potential and imprinted vorticity of supercurrent before vortex generation. The light sources are LG$_{pl,0}$. All of profiles are snapshots at $20 \tau_{GL}$.}
    \label{fig:Bz_Wn1_A2}
\end{figure*}

\begin{figure*}
    \centering
    \includegraphics[width=1\textwidth]{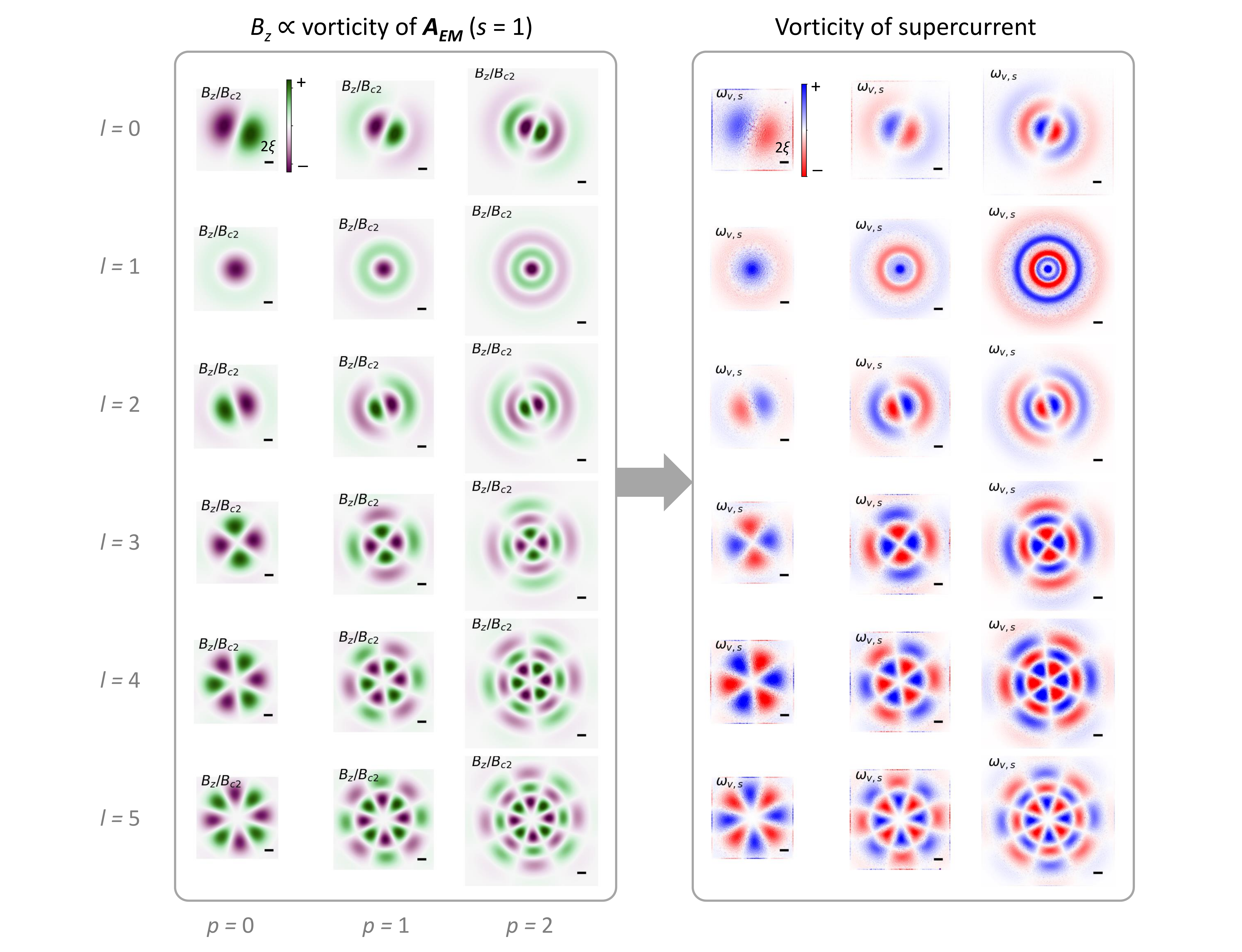}
    \caption{The vorticity of vector potential and imprinted vorticity of supercurrent before vortex generation. The light sources are LG$_{pl,1}$. All of profiles are snapshots at $20 \tau_{GL}$.}
    \label{fig:Bz_Wn1_A3}
\end{figure*}

In~\cref{sec:imprint}, the vorticity of vector potential, also equivalent to $B_z$, can directly imprint to the vorticity of supercurrent. The numerical results of imprinting profiles are demonstrated in this Appendix in Figs. ~\ref{fig:Bz_Wn1_A1}, ~\ref{fig:Bz_Wn1_A2}, and ~\ref{fig:Bz_Wn1_A3}.

\section{\label{app:shear}Similarities between~\cref{sec:sl_OAM}, we discussed the dynamics of phase branch cuts, highlighting their mechanistic similarities to trammels for the case of LG$_{01,1}$ light}. In this appendix, we demonstrate in more details the dynamics of branch cuts induced by light sources LG$_{01,0}$ and LG$_{01,1}$.

\begin{figure*}[!htbp]
    \centering
    \includegraphics[width=0.9\textwidth]{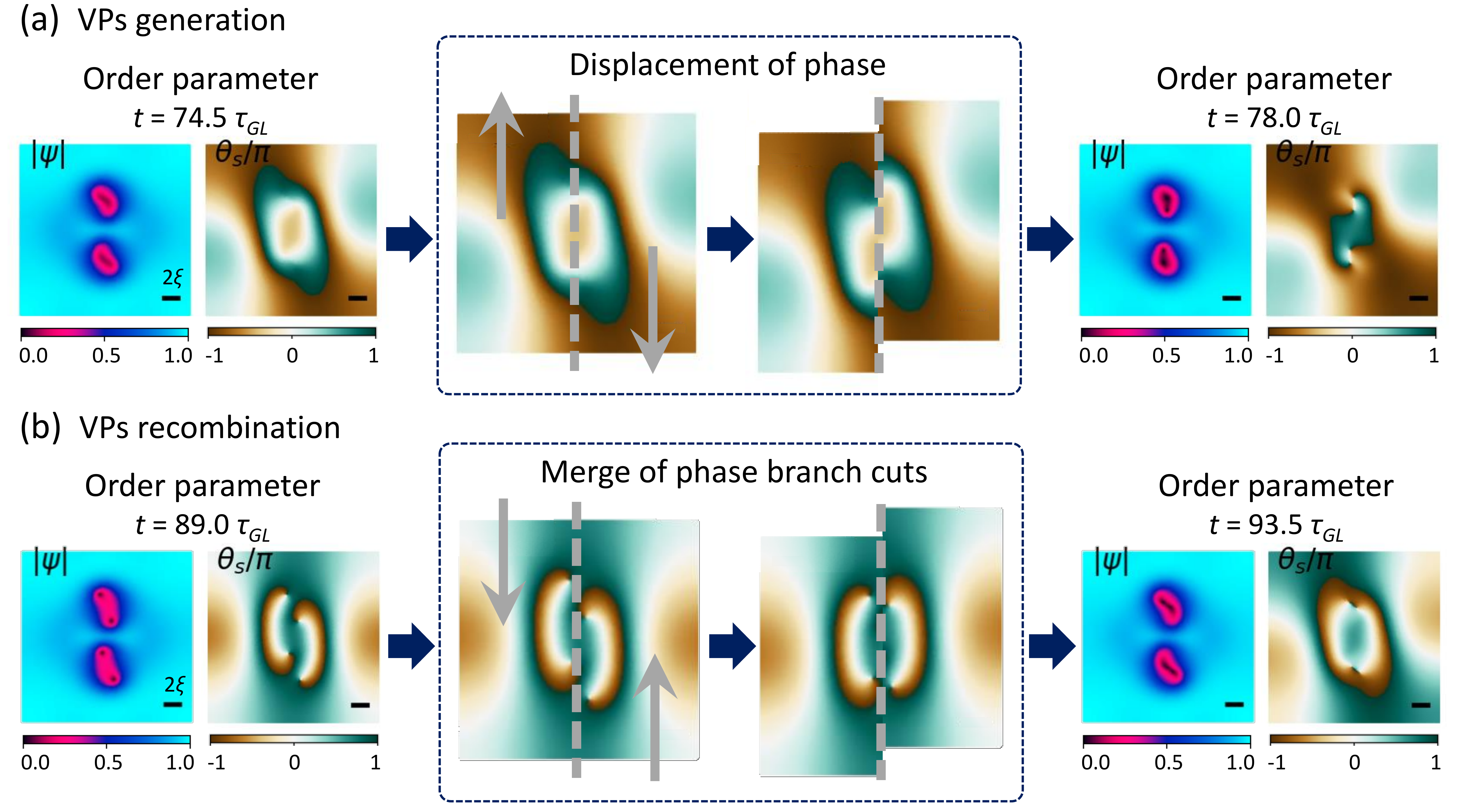}
    \caption{Demonstrations of the effects related to displacing phase induced by LG$_{01,0}$.
    (a) The VPs generation due to displacing phase. The left and right panels capture the moments just before and after VP formation, respectively, while the middle panel presents a schematic of the process. (b) VP recombination driven by the restoration of phase branch cuts. The left and right panels show snapshots of VP recombination, with the middle panel illustrating how the displacement gradually returns to its original state.
    }
    \label{fig:phase_front_shear}
\end{figure*}

For LG$_{01,0}$,~\cref{fig:phase_front_shear} exhibits  two key moments: the generation of VPs at $t=74.5-78.0 \tau_{GL}$ in (a), and the recombination VPs at $t=89.0-93.5 \tau_{GL}$ in (b). This process resembles the formation of shear displacement along $x=0$, where two VPs are generated and recombined at the endpoints of branch cuts.

\begin{figure*}[!htbp]
    \centering
    \includegraphics[width=0.9\textwidth]{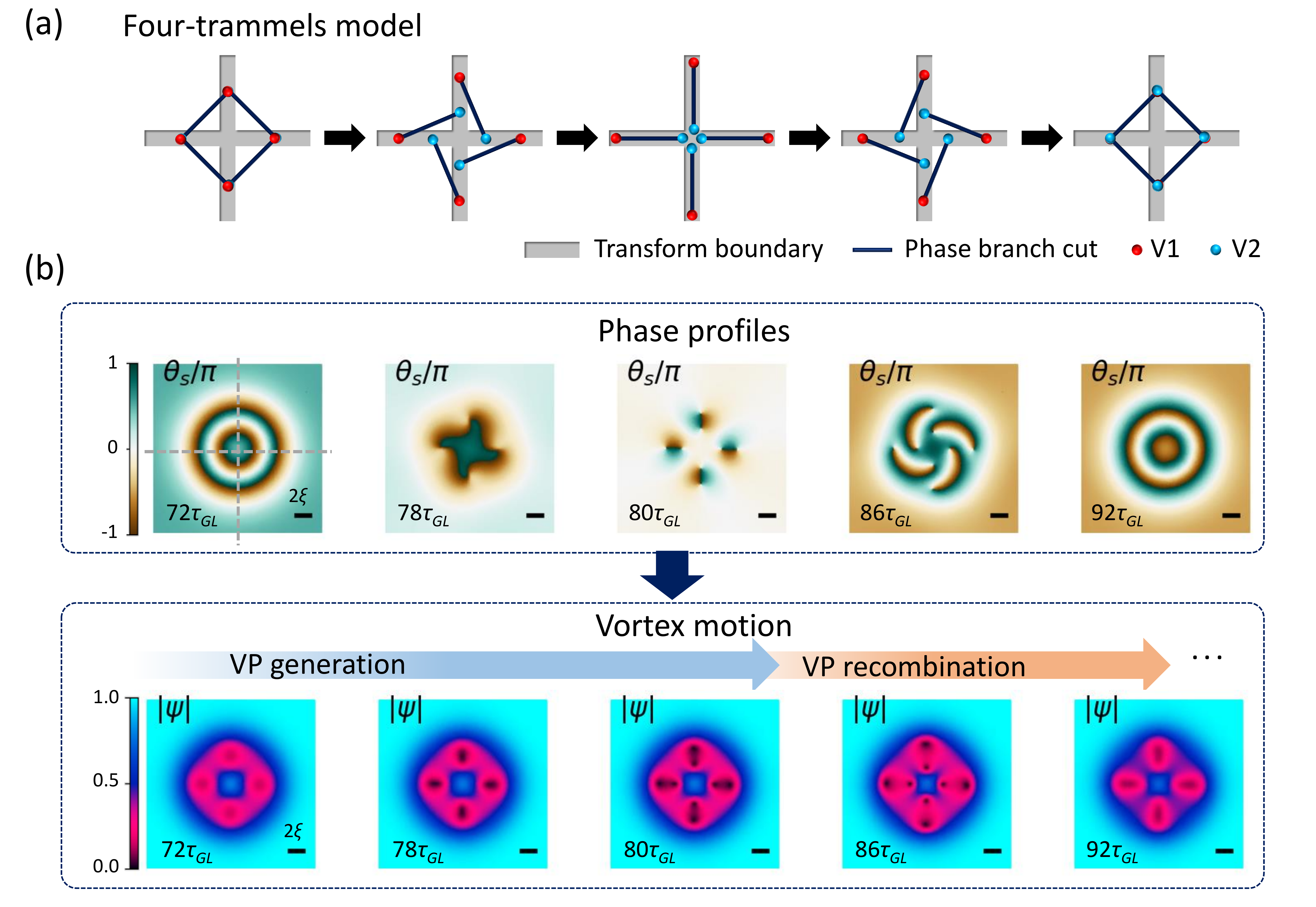}
    \caption{The trammel model induced by LG$_{01,1}$. 
    (a) Toy model of four-trammels moving along transform boundaries. The lines of boundary are shown as gray areas, and the black sticks represent the trammels in the toy model, correspond to the phase branch cuts in SC phase. The blue and red balls represent to the ends of the trammels and also indicate SC vortices with different vorticities. When the blue and red balls overlap, it signifies the annihilation of a vortex and an antivortex. (b) Snapshot of $\theta_s$ associated with the trammel model, along with the corresponding profile of $\abs{\psi}$, illustrating the vortices.}
    \label{fig:trammels}
\end{figure*}

~\cref{fig:trammels} illustrates the similarity between the trammel model and dynamics of the phase branch cuts induced by LG$_{01,1}$. The ellipsograph, a type of trammel of Archimedes~\cite{downs2003practical}, demonstrates a mechanism for converting linear motion into a circular path. In the simulation with LG$_{01,1}$, the rotational motion of phase branch cuts resembles the action of four trammel sticks as shown in~\cref{fig:trammels}, while the linear motion of VPs corresponds to the motion of the endpoints of these sticks. Recombination and generation of VPs occur when the stick endpoints overlap and detach, respectively. 
This model effectively visualizes the connection between $\theta_s$ and $\abs{\psi}$ and illustrates the underlying mechanism by which SAM and OAM are transferred to produce the linear motion of vortices.

\section{\label{app:LP}Formation of rebonding VPs with the OAM-carried light sources}

\begin{figure}[!htbp]
    \centering
    \includegraphics[width=0.5\textwidth]{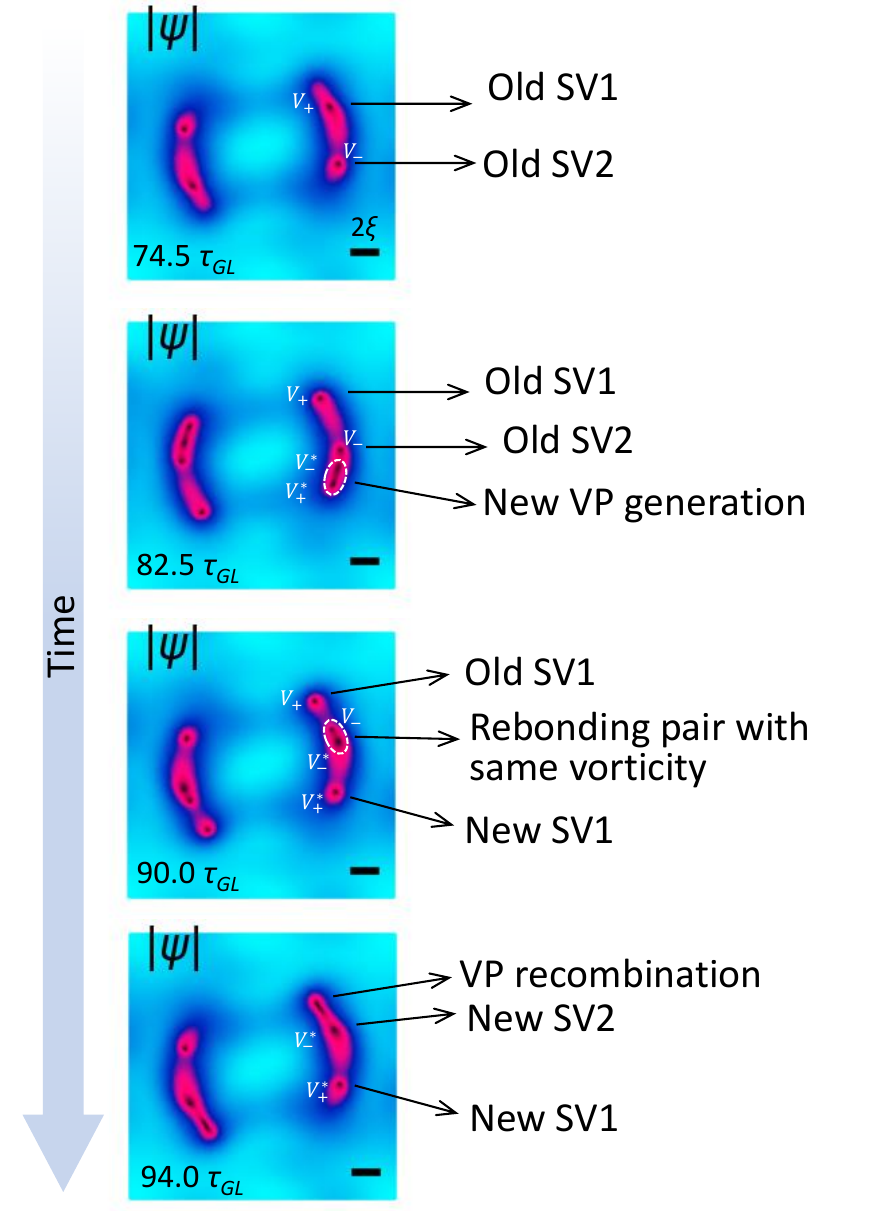}
    \caption{The time evolution of bond-unbond-rebond process of VPs with light LG$_{04,0}$. The snapshots at different time represent the four difference process. When $t=74.5 \tau_{GL}$: Old single vortex (SV) residue from the last cycle. When $t=82.5 \tau_{GL}$: New VP generation, start to debond, and rebond with the adjacent SV which has same vorticity. When $t=90. \tau_{GL}$: The rebonding pair move stably and rotate along the orbital. When $t=94.0 \tau_{GL}$: The rebonding pair bump into the old SV at the end point, and one of vortex in VP recombines with the old SV.}
    \label{fig:LP}
\end{figure}

In~\cref{sec:l>1_OAM}$d$, we mentioned that the bond-unbond-rebond process of VPs can be observed in the case of bracket-shaped orbitals. The time evolution of process is shown in ~\cref{fig:LP}. It demonstrates an observable rebonding VP of vortices with same vorticity, and residual vortices (marked SV in ~\cref{fig:LP}) at the end points of the orbital. It illustrates
both the bond-unbond-rebond process of VPs and the generation and recombination process between a residual vortex and a vortex in the rebonding VP. This sophisticated dynamics can be observed in the cases of $s=0$ and $l \geq 3$ in the supplementary material~\cite{supp1}.

\section{\label{app:psi}Snapshot of order parameters}

\begin{figure*}[!htbp]
    \centering
    \includegraphics[width=1\textwidth]{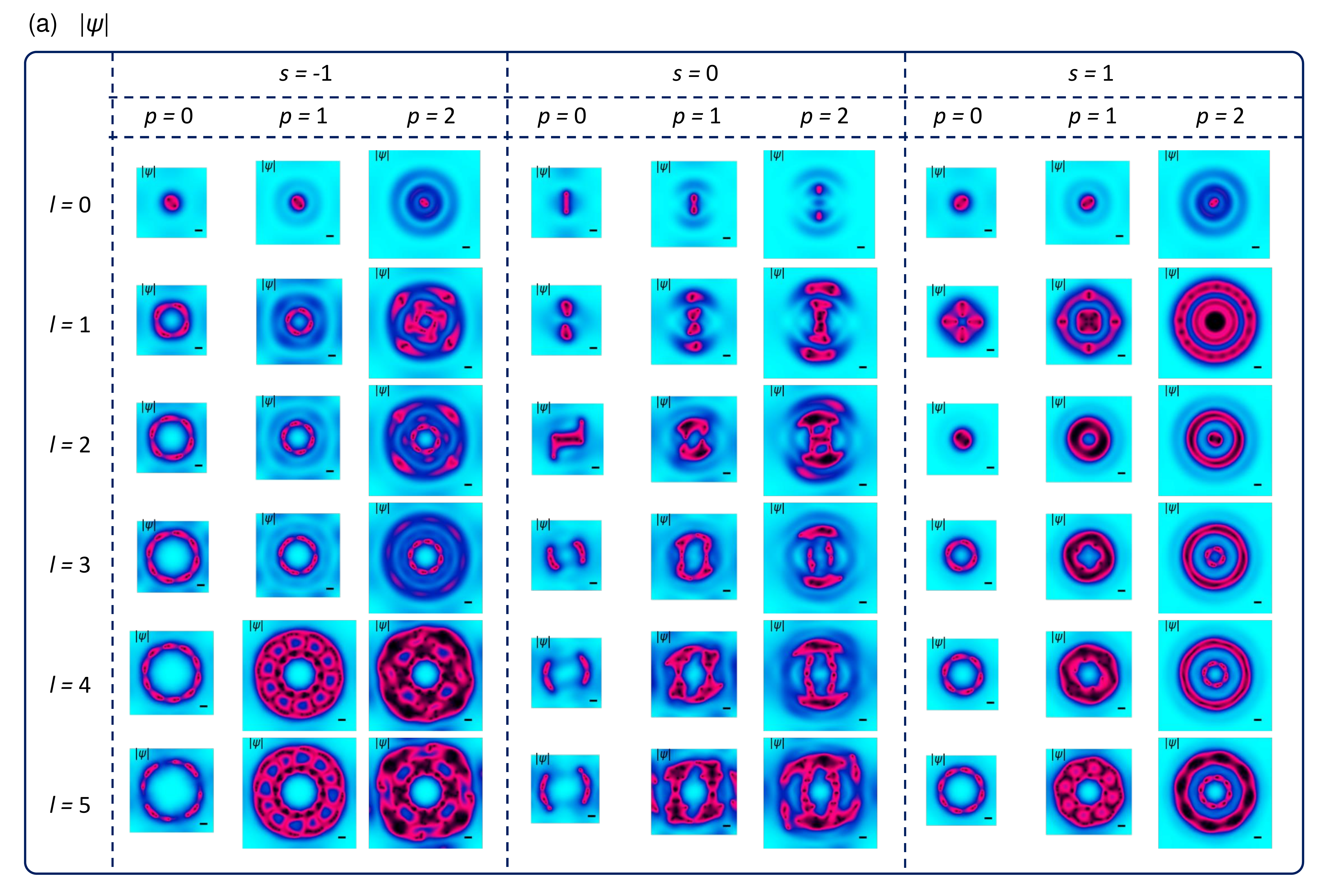}
    \includegraphics[width=1\textwidth]{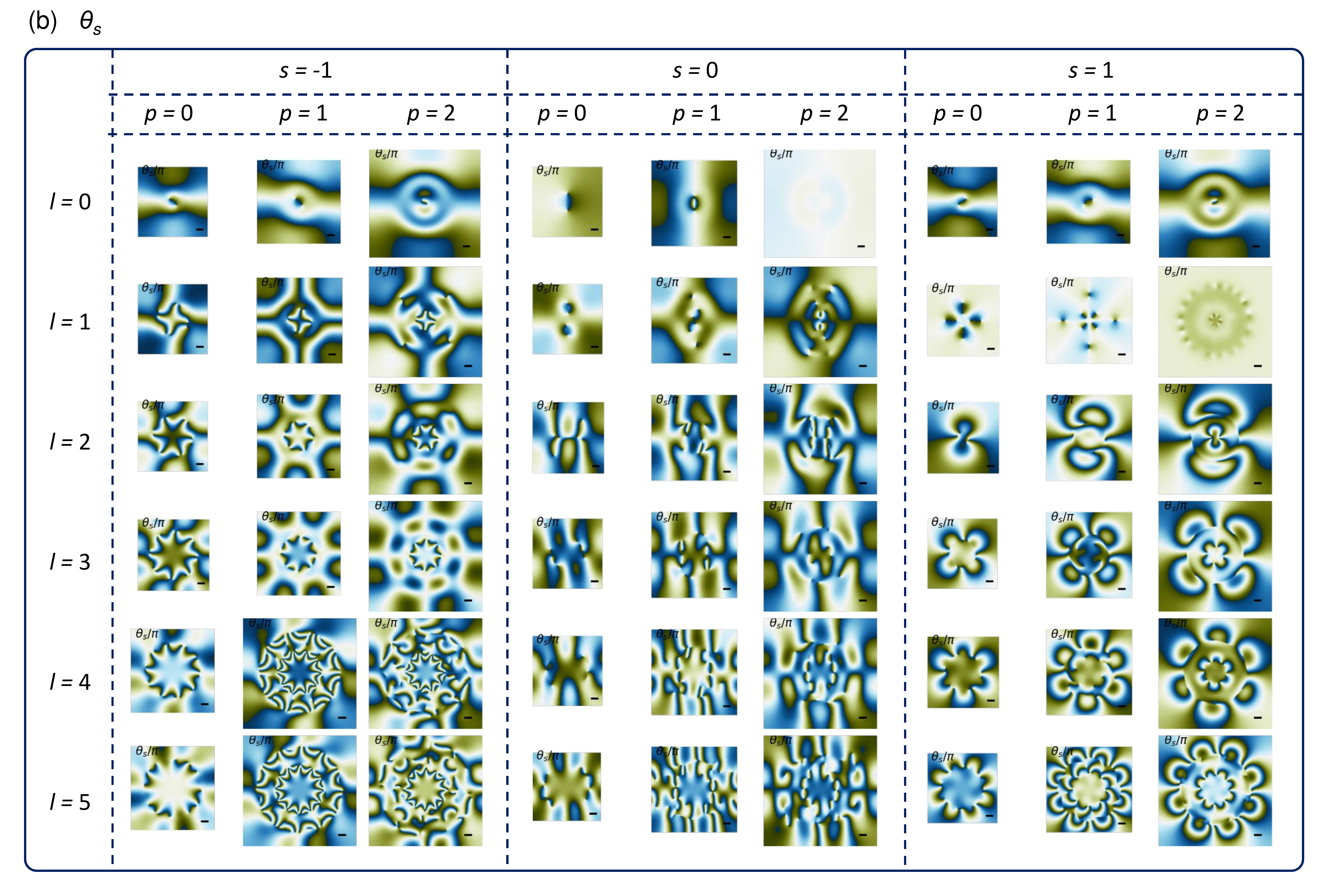}
    \caption{The snapshots of (a) $\abs{\psi}$ and (b) $\theta_s$ at $100 \tau_{GL}$.}
    \label{fig:slp}
\end{figure*}

In this Appendix, we demonstrate spatial dependence of the order parameter.
The results of simulations for $\abs{\psi}$ and $\theta_s$ for all light sources are shown in~\cref{fig:slp}(a) and (b), respectively.

\end{appendices}

\clearpage
\bibliography{LightSC2}

\begin{thebibliography}{50}
\bibitem [{sup({\natexlab{a}})}]{supp1}%
  \BibitemOpen
  \href@noop {} {} ({\natexlab{a}}),\ \bibinfo {note} {see {S}upplemental {M}aterial at URL-will-be-inserted-by-publisher for the video of light induced vortivity and vortices with light sources LG$_{pl,s}$, where $p=0, l=0\sim5, s=0$.}\BibitemShut {Stop}%
\bibitem [{sup({\natexlab{b}})}]{supp2}%
  \BibitemOpen
  \href@noop {} {} ({\natexlab{b}}),\ \bibinfo {note} {see {S}upplemental {M}aterial at URL-will-be-inserted-by-publisher for the video of light induced vortivity and vortices with light sources LG$_{pl,s}$, where $p=0, l=0\sim5, s=-1$.}\BibitemShut {Stop}%
\bibitem [{sup({\natexlab{c}})}]{supp3}%
  \BibitemOpen
  \href@noop {} {} ({\natexlab{c}}),\ \bibinfo {note} {see {S}upplemental {M}aterial at URL-will-be-inserted-by-publisher for the video of light induced vortivity and vortices with light sources LG$_{pl,s}$, where $p=0, l=0\sim5, s=1$.}\BibitemShut {Stop}%
\bibitem [{sup({\natexlab{d}})}]{supp6}%
  \BibitemOpen
  \href@noop {} {} ({\natexlab{d}}),\ \bibinfo {note} {see {S}upplemental {M}aterial at URL-will-be-inserted-by-publisher for the video of light induced vortivity and vortices with light sources LG$_{pl,s}$, where $p=1, l=0\sim5, s=1$.}\BibitemShut {Stop}%
\bibitem [{sup({\natexlab{e}})}]{supp9}%
  \BibitemOpen
  \href@noop {} {} ({\natexlab{e}}),\ \bibinfo {note} {see {S}upplemental {M}aterial at URL-will-be-inserted-by-publisher for the video of light induced vortivity and vortices with light sources LG$_{pl,s}$, where $p=2, l=0\sim5, s=1$.}\BibitemShut {Stop}%
\bibitem [{sup({\natexlab{f}})}]{supp5}%
  \BibitemOpen
  \href@noop {} {} ({\natexlab{f}}),\ \bibinfo {note} {see {S}upplemental {M}aterial at URL-will-be-inserted-by-publisher for the video of light induced vortivity and vortices with light sources LG$_{pl,s}$, where $p=1, l=0\sim5, s=-1$.}\BibitemShut {Stop}%
\bibitem [{sup({\natexlab{g}})}]{supp8}%
  \BibitemOpen
  \href@noop {} {} ({\natexlab{g}}),\ \bibinfo {note} {see {S}upplemental {M}aterial at URL-will-be-inserted-by-publisher for the video of light induced vortivity and vortices with light sources LG$_{pl,s}$, where $p=2, l=0\sim5, s=-1$.}\BibitemShut {Stop}%
\bibitem [{sup({\natexlab{h}})}]{supp4}%
  \BibitemOpen
  \href@noop {} {} ({\natexlab{h}}),\ \bibinfo {note} {see {S}upplemental {M}aterial at URL-will-be-inserted-by-publisher for the video of light induced vortivity and vortices with light sources LG$_{pl,s}$, where $p=1, l=0\sim5, s=0$.}\BibitemShut {Stop}%
\bibitem [{sup({\natexlab{i}})}]{supp7}%
  \BibitemOpen
  \href@noop {} {} ({\natexlab{i}}),\ \bibinfo {note} {see {S}upplemental {M}aterial at URL-will-be-inserted-by-publisher for the video of light induced vortivity and vortices with light sources LG$_{pl,s}$, where $p=2, l=0\sim5, s=0$.}\BibitemShut {Stop}%
\end{thebibliography}

\end{document}